\documentclass[reprint, amsmath,amssymb,aps]{revtex4-1}

\usepackage{color}

\usepackage{mathrsfs}
\usepackage{graphicx}% Include figure files
\usepackage{dcolumn}% Align table columns on decimal point
\usepackage{bm}% bold math
\usepackage{aas_macros}
\usepackage{hyperref}% add hypertext capabilities
\usepackage[mathlines]{lineno}% Enable numbering of text and display math
\begin{document}

%\preprint{APS/123-QED}

\title{Constraints on Primordial Power Spectrum from Galaxy Luminosity Functions}
%\thanks{Draft, for discussion, temporary.} %

\author{Shintaro Yoshiura}
%\email{syoshiura@unimelb.edu.au}
\affiliation{The University of Melbourne, School of Physics, Parkville, VIC 3010, Australia}

\author{Masamune Oguri}
%\email{masamune.oguri@ipmu.jp}
\affiliation{Research Center for the Early Universe, University of Tokyo, 7-3-1 Hongo, Bunkyo-ku, Tokyo 113-0033, Japan, \\
Department of Physics, University of Tokyo, 7-3-1 Hongo, Bunkyo-ku, Tokyo 113-0033, Japan, \\
Kavli Institute for the Physics and Mathematics of the Universe (Kavli IPMU, WPI), University of Tokyo, Kashiwa, Chiba 277-8583, Japan
}

\author{Keitaro Takahashi}
%\email{keitaro@kumamoto-u.ac.jp}
\affiliation{Faculty of Advanced Science and Technology, Kumamoto University, Kumamoto, Japan\\
International Research Organization for Advanced Science and Technology, Kumamoto University, Japan}

\author{Tomo Takahashi}
%\email{tomot@cc.saga-u.ac.jp}
\affiliation{Department of Physics, Saga University, Saga 840-8502, Japan}

\date{\today}% It is always \today, today,

\begin{abstract}
We derive constraints on primordial power spectrum, for the first time,  from galaxy UV luminosity functions (LFs) at high redshifts.
Since the galaxy LFs reflect an underlying halo mass function which depends on primordial fluctuations, 
one can constrain primordial power spectrum, particularly on small scales. 
We perform a Markov Chain Monte Carlo analysis  by varying parameters for primordial power spectrum
as well as those describing astrophysics. We adopt the UV LFs derived from Hubble Frontier Fields data at $z = 6 -10$, which enable us to 
probe primordial fluctuations on the scales of $k \sim 10 - 10^3~{\rm Mpc}^{-1}$.
Our analysis also clarifies how the assumption on cosmology such as primordial power spectrum affects the 
determination of astrophysical parameters.

\end{abstract}

\maketitle

%\tableofcontents

%%%%%%%%%%%%%%%%%%%%%%%%%%%%%%%%%%%%%%%%%%%%%%%%%%%%%%%%%%%%%%%%%%
%%%%%%%%%%%%%%%%%%%%%%%%%%%%%%%%%%%%%%%%%%%%%%%%%%%%%%%%%%%%%%%%%%
\section{Introduction \label{sec:intro}}

Probing primordial power spectrum is of prime importance in understanding the dynamics of the inflationary Universe. 
While observations of cosmic microwave background (CMB) have measured primordial power spectrum precisely \cite{Akrami:2018odb}, only large scales are 
probed by such observations, which gives us the information on the inflationary dynamics only within some limited period. 
To elucidate the whole inflationary dynamics, we need to probe primordial fluctuations over a much wider range of scales, which motivates the study to investigate 
constraints on primordial power spectrum with probes other than CMB, especially on small scales. 

When one assumes a standard single-field inflation model, the primordial power spectrum ${\cal P}_\zeta (k)$, with $k$ being the wave number for the mode of fluctuations, can be well described by 
the power-law form  as 
\begin{equation}
\label{eq:P_zeta_1}
{\cal P_\zeta} (k) = A_s (k_\ast) \left( \frac{k}{k_\ast} \right)^{n_s - 1},
\end{equation}
where $A_s (k_\ast)$ is the amplitude at the reference scale $k_\ast$ and $n_s$ is the spectral index. 
However, even for a standard single-field model, $n_s$ is not precisely constant, and also in some models, $n_s$ is predicted to be scale dependent,
which motivates the following description for ${\cal P}_\zeta$:
\begin{equation}
\label{eq:P_zeta_2}
{\cal P_\zeta} (k) = A_s (k_\ast) \left( \frac{k}{k_\ast} \right)^{n_s - 1 + \frac{1}{2 !} \alpha_s \ln  \left( \frac{k}{k_\ast} \right)  + \frac{1}{3!} \beta_s \ln  \left( \frac{k}{k_\ast} \right)^2},
\end{equation}
where $\alpha_s$ and $\beta_s$ are the so-called running and the running of the running parameters, respectively. 
When primordial power spectrum is probed over a wide range of scales, the running parameters can be well constrained due to 
the lever-arm effect.
Furthermore, the primordial power spectrum may not be well described by a simple form such as Eq.~\eqref{eq:P_zeta_1} or Eq.~\eqref{eq:P_zeta_2} given that a wide variety of inflationary models have been proposed, some of which do not allow those kind of smooth functional forms to describe ${\cal P}_\zeta(k)$. Therefore it would be of great importance to investigate ${\cal P}_\zeta(k)$ for a broad range of scales.
%if the potential of an inflaton is not smooth and has some ``features'', primordial power spectrum cannot be described by the above form. For instance, in multi-field inflation models, sources of primordial fluctuations can be different fields on large and small scales,and as a result the above smooth description for ${\cal P}_\zeta(k)$ is not valid anymore. One needs to probe primordial power spectrum for a wide range of scales directly to constrain such models.
These considerations motivate us to look into yet other probes of primordial fluctuations, particularly on small scales. 
Examples of such work include primordial black holes \cite{Bugaev:2008gw,Josan:2009qn,Sato-Polito:2019hws}, ultracompact minihalos \cite{Bringmann:2011ut,Emami:2017fiy}, neutral hydrogen 21cm fluctuations \cite{Kohri:2013mxa,Sekiguchi:2017cdy,Munoz:2016owz}, 21cm global signal \cite{Yoshiura:2018zts,Yoshiura:2019zxq}, CMB spectral distortion \cite{Chluba:2012we}, 
for which constraints on the small-scale amplitude of primordial power spectrum  or the so-called the running and the running of the running parameters are investigated.

In this paper, we derive constraints on primordial power spectrum using galaxy UV luminosity functions (LFs) at high redshifts for the first time.
Since the galaxy LF reflects an underlying halo mass function which depends on primordial fluctuations, one can probe primordial power 
spectrum by using observations of  UV LFs.  We adopt the LFs at high redshifts of $z = 6 - 10$ determined from Hubble Frontier Fields data \cite{Bouwens2017ApJ...843..129B,Oesch2018ApJ...855..105O,2018ApJ...854...73I}, which enable us to probe primordial power spectrum on scales much smaller than CMB,  around  $k \sim 10 - 10^3 ~{\rm Mpc}^{-1}$. 
With such small scale information, one can directly constrain its amplitude on small scales  and also the so-called running and running of the running parameters $\alpha_s$ and $\beta_s$. 

It should be noted that the LF also depends on the assumption for astrophysics such as the fraction of galactic gas converted into stars, star formation rate and so on.
To take account of such uncertainties, we perform a Markov Chain Monte Carlo (MCMC) analysis to obtain constraints on primordial power spectrum by varying both the astrophysical parameters and the ones describing
primordial power spectrum. 

We also emphasize that, although our primary purpose  is to derive constraints on primordial power spectrum,
our analysis also clarifies to what extent the determination of astrophysical parameters are affected by the assumption on primordial fluctuations.
When one tries to investigate astrophysics or constrain astrophysical parameters, the standard $\Lambda$-dominated Cold Dark Matter ($\Lambda$CDM) model with power-law primordial power spectrum is usually assumed. 
Although almost all cosmological observations are consistent with the standard $\Lambda$CDM model,
some deviations from this standard paradigm are still possible, particularly on small scales, since observations such as CMB and galaxy clustering  only probe large scales. 
In the following analysis, we model ${\cal P}_\zeta (k)$ to  capture its small scale feature keeping large scale power spectrum consistent with current observations including CMB.
This modeling enable us to investigate how the assumption on cosmology,
especially on small scales, affects the study of astrophysics.
%Our modelings of ${\cal P}_\zeta (k)$  in the following analysis to capture the features on small scales, which are yet consistent with current cosmological observations such as CMB,
%and hence it would also be helpful to investigate how the assumption on cosmology affects the study of astrophysics.

The structure of this paper is as follows. In Section~\ref{sec:method}, we give our formalism to investigate constraints on primordial power spectrum.
First we provide how we model primordial power spectrum and what parameters to be constrained. 
Then we address the formalism to derive constraints on primordial power spectrum using the galaxy luminosity function. 
In Section~\ref{sec:results}, we show constraints on ${\cal P}_\zeta(k)$ for several models from UV LFs at high redshifts. We also discuss how the uncertainties of primordial power spectrum 
affects the measurements of  astrophysical parameters in Section~\ref{sec:discussions}. 
Finally we  conclude in Section~\ref{sec:conclusion}.

%%%%%%%%%%%%%%%%%%%%%%%%%%%%%%%%%%%%%%%%%%%%%%%%%%%%%%%%%%%%%%%%%%
%%%%%%%%%%%%%%%%%%%%%%%%%%%%%%%%%%%%%%%%%%%%%%%%%%%%%%%%%%%%%%%%%%
\section{Analysis Method  \label{sec:method} }

In this section, we summarize the analysis method adopted in this paper. 
First we present our modeling of primordial power spectrum (PPS) ${\cal P}_\zeta$. We then give the formalism to 
calculate the galaxy UV LFs.  The machineries to derive constraints on PPS 
are also presented, such as the construction of the likelihood function and
a Markov Chain Monte Carlo (MCMC) analysis adopted in this paper.

%%%%%%%%%%%%%%%%%%%%%%%%%%%%%%%%%
\subsection{Modeling of Primordial Power Spectrum \label{sec:PPS}} 

We consider several types of modeling for PPS.
The first one is a power-law form which accommodates the scale dependence of
the power-law index by using the so-called running parameters:
\begin{eqnarray}
{\cal P}_{\zeta}(k) = A_s (k_\ast) \left( \frac{k}{k_*} \right)^{n_s-1  + \frac{1}{2}\alpha_s \ln\left(\frac{k}{k_*}\right) + \frac{1}{6}\beta_s\left(\ln\left(\frac{k}{k_*}\right)\right)^2},  \nonumber \\ 
\textrm{[Parametrization I]} ~~~~
\label{eq:param_I}
\end{eqnarray} 
where $n_s$ is the spectral index, $\alpha_s$ is the running parameter, and $\beta_s$ is the running of running parameter.
We adopt the reference scale of $k_*=0.05 \rm Mpc^{-1}$. 
$A_s (k_\ast) $ is the amplitude at the reference scale $k_\ast$. In our analysis, we fix the amplitude at the reference scale as $A_s (k_\ast)=2.105\times 10^{-9}$ so as to be consistent with CMB observations \cite{Akrami:2018odb}. We call this modeling as ``Parametrization I''.

Other types of modeling are based on the following form:
\begin{eqnarray}
{\cal P}_{\zeta} (k) &=& 
A_s  (k_\ast) \left(\frac{k}{k_{*}} \right)^{n_{s}-1} p(k), 
\end{eqnarray}
where $p(k)$ describes a deviation from 
the standard power-law form of ${\cal P}_\zeta$ and parametrizes 
the relative amplitude of PPS on small scales. 
In the following, we consider 3 types of functional forms for $p(k)$. 

The first one represents a step function-like form, which we denote as ``Parametrization II'',  and is given by 
\begin{eqnarray}
&& p(k)=
\begin{cases}
{p_1} \quad (10~ {\rm Mpc}^{-1}< k )\\
1~ \quad (\rm else),
\label{eq:param_II}
\end{cases}  
\notag \\
&& \hspace{40mm} \textrm{ [Parametrization II]} 
\end{eqnarray}
where $p_1$ describes the relative amplitude at the scales of $k > 10~ {\rm Mpc}^{-1}$. %This model might be replaced with more smooth function (e.g. Gaussian like function).
In this modeling, the PPS is same as the standard power-law form down to the scale of $k=10~{\rm Mpc}^{-1}$.
We choose the transition scale to be $k=10~{\rm Mpc}^{-1}$ because the galaxy UV LFs are sensitive to PPS around this scale.

Another modeling is aimed at constraining the amplitude around the scale $10~{\rm Mpc}^{-1} < k < 10^3~{\rm Mpc}^{-1}$
in more detail by separating $p(k)$ into 4 bins and parametrizing it as 
\begin{eqnarray}
p(k)&=&
\begin{cases}
{q_1} \quad (10< k <100 ~\rm Mpc^{-1})\\
{q_2} \quad (100< k <10^3 ~\rm Mpc^{-1})\\
{q_3} \quad (10^3< k ~\rm Mpc^{-1})\\
1~ \quad (\rm else) \,.
\label{eq:param_III}
\end{cases}
\notag \\ 
&& \qquad\qquad\qquad\qquad \textrm{ [Parametrization III]} 
\end{eqnarray}
Here $q_1$, $q_2$, and $q_3$ correspond to relative amplitudes on the scales denoted above. 
We take the scale smaller than $k=10^3~{\rm Mpc}^{-1}$ as one bin because, as will be shown in the next section,
our current analysis of the UV LFs is not so sensitive to scales smaller than  $k=10^3~{\rm Mpc}^{-1}$, and hence this bin is not well constrained after all.
We refer to this modeling as ``Parametrization III''.

In addition, we consider yet another modeling of $p(k)$ which  
is a step function-like one similar to Parametrization II but 
leaves the transition scale as a parameter:
\begin{eqnarray}
p(k)&=&
\begin{cases}
{r_1} \quad (k_1< k~  \rm Mpc^{-1})\\
1~ \quad (\rm else) \,,
\label{eq:param_IV}
\end{cases}
\textrm{ [Parametrization IV]} \notag \\
\end{eqnarray}
where $r_1$ is the relative amplitude at the scale of $k>k_1$. 
We refer to this modeling as 
 ``parametriztion IV''.  This parametrization enables us to probe the transition scale at which the amplitude changes.

Finally, we also analyze the case with the standard power-law form without running parameters. We consider this modeling to compare constraints on astrophysical parameters with those in other parametrizations, which allows us to investigate how the assumption on PPS affects the study of astrophysics from UV LFs.
\begin{eqnarray}
{\cal P}_{\zeta}(k) &=& A_s (k_\ast) \left( \frac{k}{k_*} \right)^{n_s-1},  
 \textrm{[Parametrization V]}
\label{eq:param_V}
\end{eqnarray} 
We call this model as ``parametriztion V''.

%%%%%%%%%%%%%%%%%%%%%%%%%%%%%%%%%
\subsection{Halo Mass Function \label{sec:mass_function}} 

Here we describe how we calculate a halo mass function (HMF) which is used to 
evaluate the galaxy luminosity function.
We basically follow the formalism  described in Ref.~\cite{2013A&C.....3...23M}.

The halo mass function is given as
\begin{eqnarray}
\frac{dn}{d\ln M} = \frac{\rho_0}{M} f(\sigma) \left| \frac{d\ln \sigma}{d \ln M}\right|,
\label{eq:dndm}
\end{eqnarray}
where $\rho_0$ is the mean matter density and
 $M$ is the mass scale corresponding to the filtering radius $R$ (see below). 
The mass variance at present time is given by
\begin{eqnarray}
\sigma_0^2(R) = \frac{1}{2\pi^2}\int_0^\infty k^2 P(k) W^2(kR) dk,
\label{eq:sigma}
\end{eqnarray}
where $P(k)$ is the linear matter power spectrum at present time and $W(kR)$ is the window function which we describe below.
The  $\sigma$ at redshift $z$ is given by applying a linear growth rate as $\sigma(M,z)=\sigma_0(M,z=0) d(z)$, where
the growth rate is
\begin{eqnarray}
d(z) = \frac{D^+(z)}{D^+(z=0)},
\end{eqnarray}
with
\begin{eqnarray}
D^+(z) = \frac{5\Omega_m}{2}\frac{H(z)}{H_0}\int_z^{\infty}\frac{(1+z')dz'}{[H(z')/H_0]^3}, 
\end{eqnarray}
and $H(z)$ is given by
\begin{eqnarray}
H(z) = H_0 \sqrt{\Omega_m(1+z)^3+\Omega_{\Lambda}},
\end{eqnarray}
where $\Omega_i$ is the density parameter for a component $i$ and $H_0$ is the Hubble constant.
From Eq.~\eqref{eq:sigma}, $d\ln \sigma / d\ln M $ in the right hand side of  Eq.~(\ref{eq:dndm}) is calculated as 
%\begin{eqnarray}
%\label{eq:dlnndm}
%\frac{d\ln \sigma}{d\ln M} = \frac{3}{2\sigma^2\pi^2R^4}\int_0^\infty \frac{dW^2(kR)}{dM}\frac{P(k)}{k^2}dk,
%\end{eqnarray}
%or alternatively one can use 
\begin{eqnarray}
\frac{d\ln \sigma}{d\ln M} &=& \frac{M}{2\sigma^2}\frac{d\sigma^2}{dM}\nonumber\\
&=&  \frac{M}{4\pi^2 \sigma_0^2} \int_0^{\infty} k^2 P(k) \frac{d W^2(kR)}{dM} dk \nonumber\\
&=&  \frac{M}{2\pi^2 \sigma_0^2} \int_0^{\infty} k^2 P(k) W(kR)\frac{d W(kR)}{dR}\frac{dR}{dM} dk. \notag \\
\label{eq:dlnsigmadlnm}
\end{eqnarray}

The function $f(\sigma)=f(\sigma(M,z))$ is usually evaluated by adopting a fitting function that is calibrated against $N$-body simulations. In what follows we adopt the formula of Sheth et al. \cite{2001MNRAS.323....1S}
\begin{eqnarray}
f(\sigma) &=& A\sqrt{\frac{2a}{\pi}}\left[ 1+ \left( \frac{\sigma^2}{a\delta_c^2}\right)^p\right]\frac{\delta_c}{\sigma}\exp{\left[- \frac{a\delta_c^2}{2\sigma^2}\right]},
\end{eqnarray}
where $A= 0.3222$, $a = 0.707$, $p=0.3$, and $\delta_c$ is 
critical density fluctuation.
%Note that the $\sigma = \sigma(M,z)$ is used for calculating the fitting formulae.

The linear matter power spectrum at present time is related with primordial power spectrum ${\cal P}_{\zeta} (k)$ as
\begin{eqnarray}
P(k) \propto {\cal P}_{\zeta} (k) T(k)^2,
% P(k) = (2\pi^2 k){\cal P}_{\zeta} (k) T(k)^2,
\end{eqnarray}
where $T(k)$ is the transfer function which can be derived by linear Boltzmann solvers such as CAMB \cite{Lewis:1999bs} or by appropriate fitting functions (e.g., \cite{1999ApJ...511....5E}). Here we use python CAMB \footnote{https://camb.readthedocs.io} to calculate the transfer function for a given cosmological model. Since we are interested only in cosmological models that are compatible with CMB observations, the amplitude of PPS at the reference scale $A_s (k_\ast=0.05)=2.105\times 10^{-9}$ is fixed in this work. In the following analysis, other cosmological parameters that are tightly constrained by CMB observations are also fixed as $\Omega_mh^2 = 0.142$, $\Omega_bh^2=0.0224$, $\Omega_\Lambda=1-\Omega_m$, and $H_0=67.7~{\rm km}/{\rm s}/{\rm Mpc}$, where  $h$ is the normalized Hubble constant in units of $100 {\rm km}/{\rm s}/{\rm Mpc}$. This setup, especially fixing of the amplitude of PPS at the reference scale, keeps the fit to CMB data almost unchanged for our models of PPS.

Regarding the window function, we consider the following three filters:  (i)~top-hat filter, (ii)~Gaussian filter, and (iii)~smooth-$k$ filter.
The top-hat window function is given by 
\begin{eqnarray}
W(kR) = \frac{3[\sin(kR)-kR\cos(kR)]}{(kR)^3}.
\end{eqnarray}
The mass $M$ is expressed as
\begin{eqnarray}
M = \frac{4\pi \rho_0}{3} R^3,
\label{eq:mass_scale}
\end{eqnarray}
where $R$ is radius corresponding to the mass scale.
The derivatives of $W(kR)$ and $M$ with respect to $R$ are given by
\begin{eqnarray}
\frac{dW(kR)}{dR} &=& \frac{3}{kR^2}\left[ \sin(kR)\left(1 - 3 \frac{1}{k^2R^2}\right) + 3\frac{\cos(kR)}{kR} \right] \,, \notag \\ \\
\frac{dM}{dR} &=& 4\pi \rho_0 R^2.
\end{eqnarray}

%\begin{eqnarray}
%\frac{dW^2}{dM} &=& [\sin(kR)-kR\cos(kR)] \nonumber\\
%&\times& \left[\sin(kR) \left(1-\frac{3}{(kR)^2} \right) + 3\frac{\cos(kR)}{kR} \right].
%\end{eqnarray}

For the Gaussian filter, we adopt the following forms for the window function and the mass
\begin{eqnarray}
W(kR) &=& \exp\left(-\frac{k^2R^2}{2}\right), \\
\label{eq:mass_scale_gaussian}
M(R) &=& (2\pi)^{1.5}\rho_0 R^3.
\end{eqnarray}
The derivatives are given by 
\begin{eqnarray}
\frac{dW(kR)}{dR} &=& -k^2R \exp{\left(-\frac{k^2R^2}{2}\right)},\\
\frac{dM}{dR} &=&  3(2\pi)^{1.5}\rho_0 R^2.
\end{eqnarray}

Although the top-hat and the Gaussian filters are commonly used in the literature, 
we also consider the so-called smooth-$k$ filter \cite{Leo:2018odn} which has the following form
\begin{eqnarray}
W(kR) &=& \frac{1}{1+(kR)^\beta},  \\
\frac{dW(kR)}{dR} &=& - \frac{\beta k (kR^{\beta-1})}{(1+(kR)^{\beta})^2}.
\end{eqnarray}
The mass is given by 
\begin{eqnarray}
\label{eq:mass_scale_smoothk}
M(R) &=& \frac{4\pi \rho_0}{3} (cR)^3,
\end{eqnarray}
where $\beta$  and $c$ are free parameters in this filter. Below we fix these parameters as $\beta= 2.0$ and $c =3.15$ so that the mass function  becomes the same as that for the top-hat filter with the reference model at $z=0$ where the standard power-law form is assumed for PPS (Parametriztion V) and cosmological parameters given above with $n_s = 0.968$ are adopted.
We consider this filter function because it gives a good fit to $N$-body simulations for models with suppressed matter power spectrum such as the warm dark matter case \cite{Leo:2018odn}, which resembles of our analysis of PPS with enhancements or suppressions on small scales. Although the sharp-$k$ filter is another common choice of a filter (e.g., \cite{2015MNRAS.451.3117S}), we do not use this filter in this work because the smooth-k filter is introduced in \cite{Leo:2018odn} as 
an updated form of the sharp-$k$ filter to reproduce $N$-body simulation results for power spectra truncated at small mass scales.
We consider these three filters to take account of the uncertainty associated with the choice of the filter. Ultimately such choice should be checked and validated with $N$-body simulations with features of PPS as considered in this paper, which is beyond the scope of this paper.
%Indeed we need to check whether this filter is appropriate for our case, however, this is beyond the scope of this paper, and by investigating constraints adopting 
%three types of filter mentioned above, we can see to what extent the filter affects our results. 
%We also discuss this issue in the next section.

\begin{figure}[h]
\includegraphics[width=7.5cm]{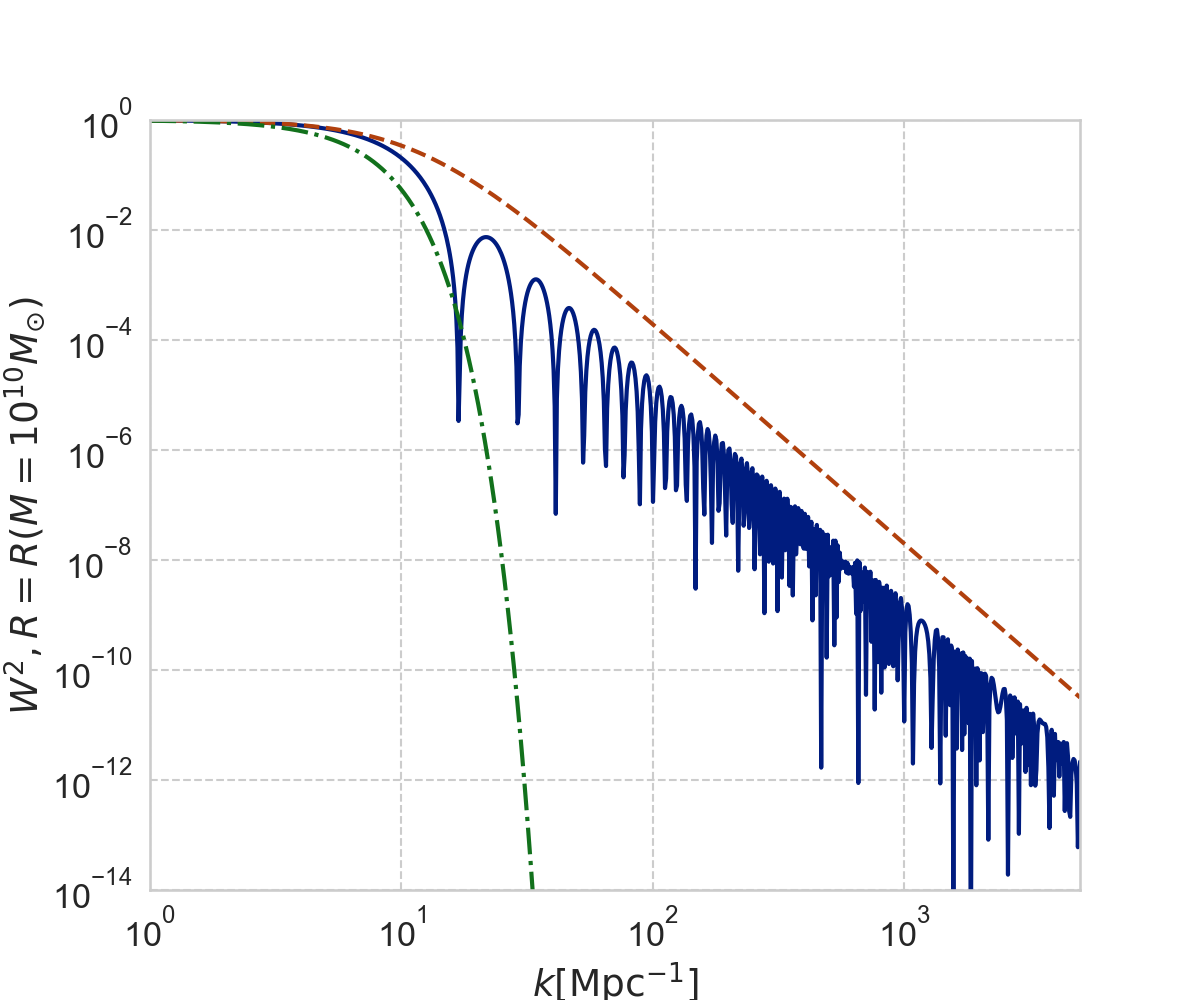}
\caption{\label{fig:MUV2Mhalo2} 
Comparison of window functions $W^2$ for the mass scale $M=10^{10} M_\odot$. The solid line is the top-hat filter, the dot-dashed line is the Gaussian filter, and the dashed line is the smooth-$k$ filter with $(\beta,c) = (2.0, 3.15)$. }
\end{figure}

In Fig.~\ref{fig:MUV2Mhalo2}, we show the window functions of top-hat, Gaussian, and smooth-$k$ filter with $(\beta,c) = (2.0, 3.15)$. The top-hat filter has oscillating feature around the tail, and the oscillation can make artificial shape in HMF for enhanced PPS models, as has been found in \citep{Yoshiura:2019zxq}. The smooth-$k$ filter, on the other hand, has smooth response to the enhancement of PPS at small scales. 
%Since appropriate filter is unknown, we also make an analysis by using the Gaussian filter for comparison. 
The Gaussian filter erases the contribution from PPS at small scales, leading to the LFs at $z=6$ roughly 1.3 times smaller than that with the top-hat filter. 

%%%%%%%%%%%%%%%%%%%%%%%%%%%%%%%%%
\subsection{Luminosity Function \label{sec:luminosity_func}} 

Once the halo mass function is calculated, we can evaluate the UV luminosity function as follows \cite{Park2019MNRAS.484..933P}. We refer the readers to \cite{Park2019MNRAS.484..933P} and references therein for the details of the analysis. 

First, the stellar mass of a galaxy is converted from halo mass $M_{\rm h}=M$ as 
\begin{eqnarray}
M_*(M_{\rm h}) = f_{\rm star} \left(\frac{\Omega_b}{\Omega_m}\right)M_{\rm h},
\end{eqnarray}
where $f_{\rm star}\le 1$ is the fraction of galactic gas in stars.
This fraction is modeled as 
\begin{eqnarray}
f_{\rm star}(M_{\rm h}) = f_{*,10}\left( \frac{M_{\rm h}}{10^{10}M_{\odot}}\right)^{\alpha_{\rm star}},
\end{eqnarray}
where $f_{*,10}$ is fraction of galactic gas in stars
for  $M_{\rm h}=10^{10} M_{\odot}$ and the power law from is assumed with $\alpha_{\rm star}$ being a power-law index.

Next, the star formation rate (SFR) is assumed as
\begin{eqnarray}
\dot{M}_* (M_{\rm h},z) = \frac{M_* (M_h)}{t_{\rm star}H^{-1}(z)},
\label{eq:SFR_Mhalo}
\end{eqnarray}
where $t_{\rm star}$ controls the star formation rate in unit of Hubble time. The range is restricted to $0<t_{\rm star}<1$. 
The SFR is related to the rest-frame UV luminosity, which is written as
\begin{eqnarray}
\dot{M}_*(M_{\rm h},z) = \mathscr{K}_{\rm UV} \times L_{\rm UV},
\label{eq:SFR_LUV}
\end{eqnarray}
where $\mathscr{K}_{\rm UV} =1.15\times 10^{-28} \rm M_{\odot} yr^{-1} /ergs \,s^{-1}\, Hz^{-1}$ \cite{2016MNRAS.460..417S}.
The UV luminosity is converted to UV magnitude using the AB magnitude relation \cite{1983ApJ...266..713O},
\begin{eqnarray}
\log_{10}\left(\frac{L_{\rm UV}}{\rm ergs^{-1}Hz^{-1}}\right) = 0.4 \times(51.63-M_{\rm UV}).
\label{eq:UVLF_MUV}
\end{eqnarray}

The SFR in a small mass halo might be suppressed by feedback, which is modeled as
\begin{eqnarray}
f_{\rm duty}(M_{\rm h}) = \exp\left(-\frac{M_{\rm turn}}{M_{\rm h}}\right),
\end{eqnarray}
where $M_{\rm turn}$ is the mass threshold defined such that a halo with the mass below $M_{\rm turn}$ cannot form stars efficiently.

Combining these models, the UV luminosity function is given by
\begin{eqnarray}
\phi (M_{\rm UV}) = f_{\rm duty}\frac{dn}{dM_{\rm h}}\left| \frac{dM_{\rm h}}{dM_{\rm UV}} \right|, 
\end{eqnarray}
where $M_{\rm UV}$ is the UV magnitude which is related to the UV luminosity as given by Eq.~\eqref{eq:UVLF_MUV}.

In addition to the UV LFs, we can also utilize the stellar to halo mass ratio (SHMR) at high redshifts, which has been reported in \cite{2016ApJ...821..123H}.  We also include this observation to constrain the astrophysical parameters. %We use Eqs (58) to (60) and Table. 5 and Table. 6 of  \cite{2016ApJ...821..123H}.

%%%%%%%%%%%%%%%%%%%%%%%%%%%%%%%%%
\subsection{MCMC analysis and Likelihood \label{sec:MCMC_like}}

Although our primary purpose is to derive constraints on PPS, we also need to take account of astrophysical parameters since they also 
affect the UV LFs. We thus perform a Markov Chain Monte Carlo (MCMC) analysis to explore the likelihood as a function of many model parameters. 
Parameters varied in our analysis and their ranges are summarized in Table~\ref{tab:range}. Since astrophysical parameters such as $f_{*,10}$, $\alpha_{\rm star}$, $M_{\rm turn}$, and $t_{\rm star} $ can evolve in time, to be conservative, we treat these astrophysical parameters independently for 5 different redshift bins. Hence we vary 20  parameters in total for astrophysical ones. Besides these parameters, there are also several PPS parameters, where the total number of the model parameters depends on the parametrization.
The MCMC analysis is performed using the public code {\tt emcee} \cite{emcee}. 
Our criterion for the convergence of the chain is based on the integrated auto-correlation time $\tau$  \cite{2010CAMCS...5...65G}. We accept chain longer than 50$\tau$ and check the convergence such that the value of $\tau$ is almost unchanged even if we increase the number of samples as suggested in \cite{emcee}.

\begin{table}[h]
\caption{\label{tab:range}%
Parameters and their prior ranges in our MCMC analyais.
We assume a flat prior within the range for all parameters.}
\begin{ruledtabular}
\begin{tabular}{ccccc}
param  &min &max & Prior\\
%\mbox{Three}&\mbox{Four}&\mbox{Five}\\
\hline
$f_{*,10} ~(z=6,7,8,9, 10)$  & 0.001 & 1.0 & log\\
$\alpha_{\rm star} ~(z=6,7,8,9, 10)$ & -0.5 & 1.0 & linear\\
$M_{\rm turn}~(z=6,7,8,9, 10)$ &  $10^7 M_\odot$ & $10^{10} M_\odot$ & log\\
$t_{\rm star}~(z=6,7,8,9, 10)$  & 0.0 & 1.0 & linear\\
$n_s$ & 0.0 & 2.0 & linear\\
$\alpha_s$  & -1 & 1 & linear\\
$\beta_s$ & -1 & 1 & linear\\
$p_1$ &  $10^{-5}$ & $10^8$ & log\\
$q_1$ &  $10^{-5}$ & $10^8$ & log  \\
$q_2$ &  $10^{-5}$ & $10^8$ & log  \\
$q_3$ & $10^{-5}$ & $10^8$ & log  \\
$r_1$ &  $10^{-5}$ & $10^8$ & log  \\
$k_1$ &  10.0 & $10^4$ & log\\
\end{tabular}
\end{ruledtabular}
\end{table}
% \begin{table}[h]
% \caption{\label{tab:range}%
% Parameters varied and their ranges in our MCMC analyais.
% We assume a flat prior for all parameters.}
% \begin{ruledtabular}
% \begin{tabular}{ccccc}
% param &fid &min &max &\\
% %\mbox{Three}&\mbox{Four}&\mbox{Five}\\
% \hline
% $f_{*,10}$ & 0.05, & 0.001 & 1.0 & log\\
% $\alpha_{\rm star}$ & 0.5, & -0.5 & 1.0 & liner\\
% $M_{\rm turn}$ & $5\times10^8$ & $10^7$ & $10^{10}$ & log\\
% $t_{\rm star}$ & 0.5, & 0.0 & 1.0 & liner\\
% $n_s$ & 0.968, & 0.0 & 2.0 & liner\\
% $\alpha_s$ & 0.0, & -1 & 1 & liner\\
% $\beta_s$ & 0.0, & -1 & 1 & liner\\
% $p_1$ & 1.0 & $10^{-5}$ & $10^8$ & log\\
% $q_1$ & 1.0 & $10^{-5}$ & $10^8$ & log  \\
% $q_2$ & 1.0 & $10^{-5}$ & $10^8$ & log  \\
% $q_3$ & 1.0 & $10^{-5}$ & $10^8$ & log  \\
% $r_1$ & 1.0 & $10^{-5}$ & $10^8$ & log  \\
% $k_1$ & 10.0 & 10.0 & $10^4$ & log\\
% \end{tabular}
% \end{ruledtabular}
% \end{table}

\begin{figure*}[!t]
\includegraphics[width=20.0cm]{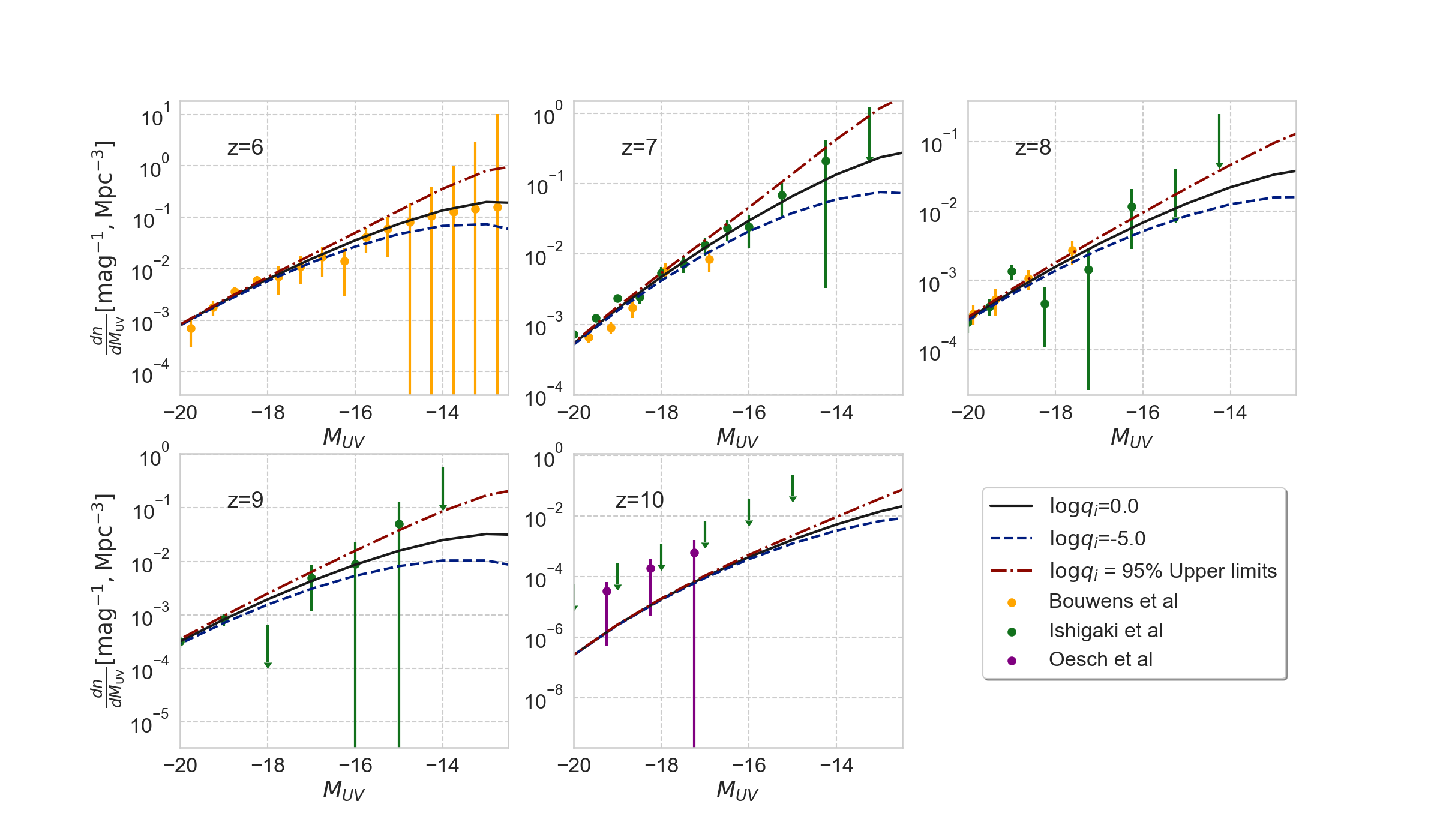}
\caption{\label{fig:LFall} LFs adopted in our analysis. We use those at $z=6$ \cite{Bouwens2017ApJ...843..129B}, $z=7$ \cite{Bouwens2017ApJ...843..129B,2018ApJ...854...73I}, $z=8$ \cite{2018ApJ...854...73I}, $z=9$ \cite{2018ApJ...854...73I} and $z=10$ \cite{Oesch2018ApJ...855..105O,2018ApJ...854...73I}. Solid, dashed and  dot-dashed lines are the LFs for Parametrization III with $\log q_i=0$, $\log q_i=-5$, and $q_i$ values of 95\% upper limit for all $i$. For astrophysical parameters, we use the best-fit parameters obtained in the MCMC for Parametrization III. }
\end{figure*}

In our MCMC analysis, the likilihood of each model is evaluated by comparing the model predcited UV LFs at high redshifts $z=6 -10$ with those derived from Hubble Frontier Fields data, which is depicted in Fig.~\ref{fig:LFall}. 
More specifically, we assume the following log likelihood for the LFs 
\begin{eqnarray}
\ln \mathcal{L}_{\rm LF1} =  - \frac{1}{2}\sum_i {\frac{(y_i-m_i)^2}{\sigma_i^2}},
\end{eqnarray}
where $y_i$ is the observed UV LF, $m_i$ is the LF calculated based the formalism given in the previous section for a given PPS Parametrization and model parameters, and $\sigma_i$ is the error of the observation. The label $i$ refers to each data depicted in Fig.~\ref{fig:LFall}. For simplicity, if the asymmetric error is given, we use its averaged value. 
%Since the 1st term in the right hand side just gives a constant contribution to the likelihood, we ignore it in our analysis.
 
When only an upper limit is given in the observed LFs data, 
the log likelihood is calculated by assuming a Poisson distribution with sample 0 as
\begin{eqnarray}
\ln \mathcal{L}_{\rm LF2} = -\sum_i \frac{m_i}{(y_i/1.15)},
\end{eqnarray}
where 1.15 indicates the expected number of galaxies in the bin corresponding to 1$\sigma$ upper limit assuming the Poisson statistics.
%where $\lambda$ is the average value of the distribution. %For instance the 1 $\sigma$ (68\%) upper limit as z=7 and at $M_{\rm UV}=-12$ is $10^{0.085}$ which corresponds to $\exp{1.15}/V_{\rm eff}$?? Thus, $\lambda=y*V_{\rm eff}$.

The parameters can be constrained using the LFs for each redshift and combining all redshift data. We give our main results from the combination of  all redshift data in the next section.   Note that the LF of bright galaxies can be suppressed by AGN feedback which is not considered in this model, and therefore we use the LF only at $M_{\rm UV}>-20$, which is the same treatment  as in \cite{Park2019MNRAS.484..933P}.

%We use the smooth-$k$ filter in for several PPS models as a reference, but cases for the top-hat window and Gaussian window are also presented for comparison. Note that the most appropriate filter should be chosen by comparing with the massive numerical N-body simulation. 

Based on \cite{2016ApJ...821..123H}, we can also utilize the observed stellar to halo mass ratio (SHMR) to improve constraints. The SHMR is derived from clustering analysis using Hubble deep image and Subaru Hyper Suprime-Cam data. Its likelihood is calculated as
\begin{eqnarray}
\ln \mathcal{L}_{\rm SHMR} = -\frac12 \sum_i \frac{ (M_{{\rm halo},i}(M_{\rm UV}) - M_{{\rm halo},i,{\rm obs}}(M_{\rm UV}))^2}{{\sigma_{{\rm SHMR},i}^2}}, \notag \\
\end{eqnarray}
where $M_{\rm halo,obs}$ is the observed halo mass for a given $M_{\rm UV}$ and $\sigma_{\rm SHMR}$ is its error. By adding the data from SHMR, the constraints on the astrophysical parameters are expected to improve. Since the data is provided in the redshift range of $z=4 - 7$, we use those for $z=6$ in our MCMC analysis. The inclusion of this SHMR data will be discussed in Section~\ref{sec:SHMR}.

Since the form of PPS on large scales has been well constrained by CMB data, in addition to the data from UV LFs and SHMR, we add the likelihood function which describes the constraint on the spectral index $n_s$ from Planck, which effectively combine our analysis with large scale observations of CMB, considering the fact that cosmological parameter sets that are tightly constrained by CMB observations are fixed in our analysis. 
This can be done by adding the likelihood function for $n_s$
\begin{eqnarray}
\ln \mathcal{L}_{\rm Planck} =
- \frac{1}{2}  \frac{(n_s - n_{s(\rm planck)})^2}{\sigma^2_{n_s}},
\end{eqnarray}
where we take $n_{s \rm (Planck)} = 0.9665$ and $\sigma_{n_s} = 0.0038$ \cite{Akrami:2018odb}.

The total likelihood function is given by the sum of $\ln {\cal L}$ discussed above as
\begin{equation}
\ln {\cal L}_{\rm total} 
= \ln \mathcal{L}_{\rm LF1} + \ln \mathcal{L}_{\rm LF2}
+ \ln \mathcal{L}_{\rm SHMR} + \ln \mathcal{L}_{\rm Planck},
\end{equation}
with which we investigate constraints on the PPS parameters  as well as the astrophysical ones for each redshift.

%%%%%%%%%%%%%%%%%%%%%%%%%%%%%%%%%%%%%%%%%%%%%%%%%%%%%%%%%%%%%%%%%%
%%%%%%%%%%%%%%%%%%%%%%%%%%%%%%%%%%%%%%%%%%%%%%%%%%%%%%%%%%%%%%%%%%
\section{Results \label{sec:results}} 

Now we present our constraints on the PPS for each Parametrization. We here focus only on key results. The full MCMC corner plots are shown in Appendix~\ref{sec:Afr}. We also discuss the effect of the stellar to halo mass ratio on the constraints. Although the main purpose of this paper is to derive constraints on the PPS, we also investigate how the cosmological assumption such as the form of PPS affects constraints on astrophysical parameters.

%%%%%%%%%%%%%%%%%%%%%%%%%%%%%%%%%
\subsection{\label{sec:const_param_I} Constraints on PPS: Parametrization I (runnings)} 

First we present our results for the case of Parametrization I where the PPS is characterized by the spectral index $n_s$, the running $\alpha_s$, and the running of the running $\beta_s$. 

%In this case, we constraint the parameters using LFs at all redshifts simultaneously. Then, the astrophysical parameter should be independent at different redshifts. Then, there are 5 redshift bin ($z$=6, 7, 8, 9 and 10), 4 astrophysical parameter, spectral index $n_s$ and 2 running parameters.

\begin{figure}[h]
\includegraphics[width=7.5cm]{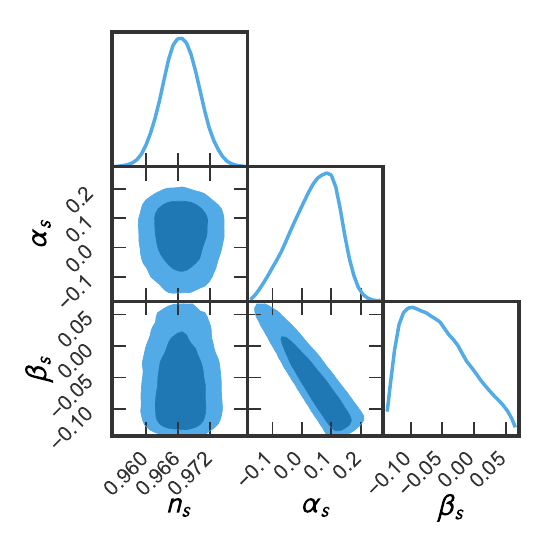}
\caption{\label{fig:nab} Constraints on $n_s$, $\alpha_s$, and $\beta_s$ for Parametrization I.
68 \% and 95\% allowed regions are depicted as dark and light blue, respectively. Astrophysical parameters are marginalized. Here the top-hat filter is assumed.  }
\end{figure}

Constraints obtained from the UV LFs and assuming the top-hat filter are
\begin{eqnarray}
\label{eq:abz6}
&& n_s = 0.967_{-0.004}^{+0.003},\nonumber\\ [5pt]
&& \alpha_s = 0.050_{-0.090}^{+0.070},\nonumber\\ [5pt]
&& \beta_s = -0.063_{-0.046}^{+0.061}.
\end{eqnarray}
The corner plot is shown in Fig.~\ref{fig:nab}.
When we adopt the smooth-$k$ filter, constraints on 
these parameters are given as
\begin{eqnarray}
\label{eq:abz6_smoothk}
&& n_s = 0.967_{-0.004}^{+0.003},\nonumber\\ [5pt]
&& \alpha_s = 0.046_{-0.099}^{+0.077},\nonumber\\ [5pt]
&& \beta_s = -0.057_{-0.049}^{+0.062}.
\end{eqnarray}

\begin{figure}[h]
\includegraphics[width=10.0cm]{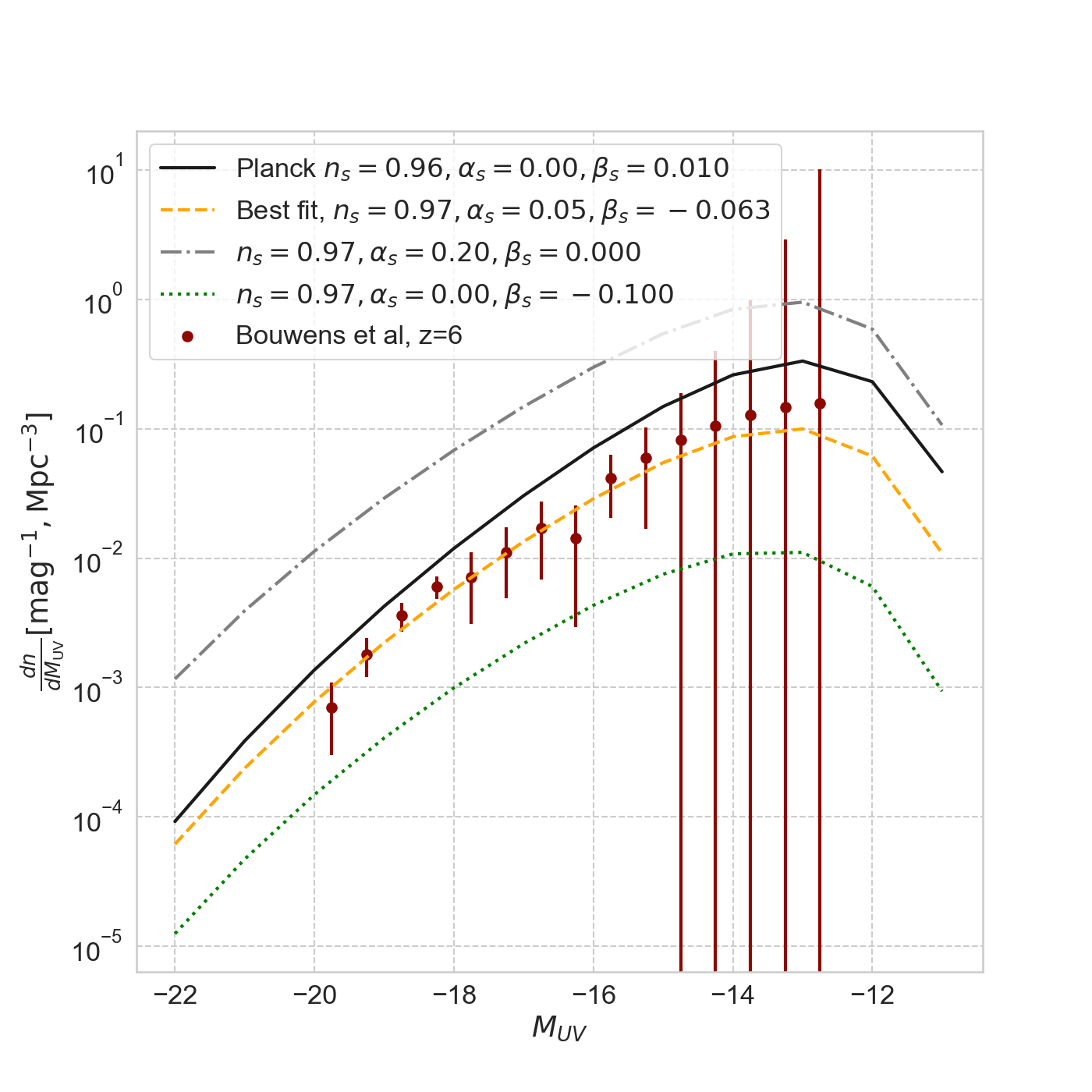}
\caption{\label{fig:nabLF} Comparison of the observed and model predicted LFs at $z=6$ for Parametrization I. Solid and dashed lines are the LFs with  best-fit running parameters from Planck and the best fit value of our results with the top-hat filter (Eq.~(\ref{eq:abz6})), respectively.  For reference, two extreme models are shown as dot-dashed line ($n_s = 0.97$,  $\alpha_s=0.2$, and $\beta_s=0.0$) and dotted line ($n_s = 0.97$,  $\alpha_s=0.0$, and $\beta_s=-0.1$). Astrophysical parameters are taken 
to be the best-fit values, $\log f_{*,10} = -0.91$, 
$\alpha_{\rm star}  = 0.28$,  $\log M_{\rm turn}= 8.9$, and 
$t_{\rm star}=  0.55$.}
\end{figure}

The constraints above can be compared with the ones from Planck 2018 (Eqs.~(16)-(18) in \cite{Akrami:2018odb}):
\begin{eqnarray} 
&& n_s = 0.9625\pm0.0048,  \nonumber\\ [5pt]
&& \alpha_s = 0.002 \pm {0.010}, \nonumber\\ [5pt]
&& \beta_s = 0.010 \pm {0.013}.
\end{eqnarray}
Our constraints from the UV LFs are consistent with those from Planck. The LF at $z=6$ assuming the mean values given in Eq.~\eqref{eq:abz6} are shown in Fig.~\ref{fig:nabLF}. For reference, we also plot the cases with $(n_s, \alpha_{s}, \beta_s)=(0.97, 0.2, 0.0)$, and $(n_s, \alpha_{s},\beta_s) =(0.97, 0.0, -0.1)$. 
From Fig.~\ref{fig:nabLF}, one can see that these two cases are completely deviated from the data. Indeed these two models are  excluded at 95 \% CL as shown in Fig.~\ref{fig:nab}.
\begin{figure*}[!t]
\includegraphics[width=3.5cm]{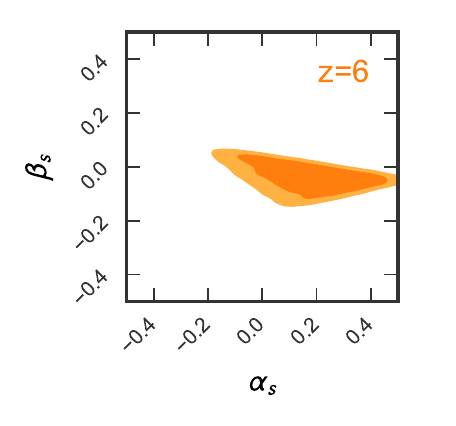}
\includegraphics[width=3.5cm]{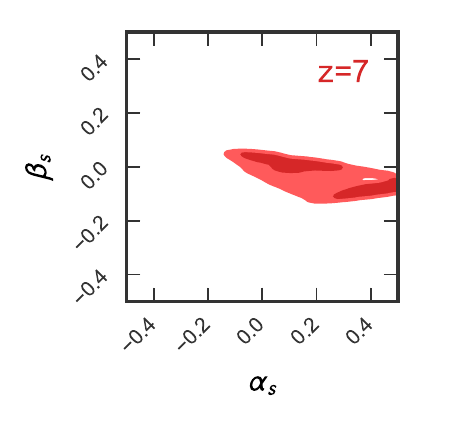}
\includegraphics[width=3.5cm]{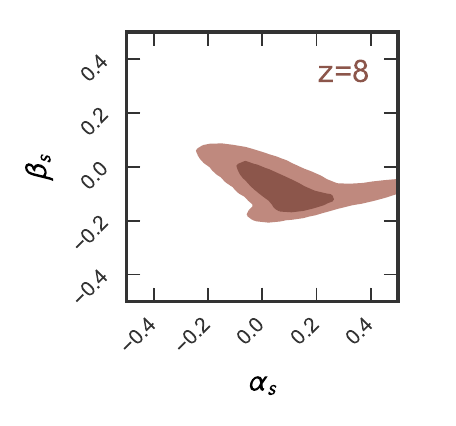}
\includegraphics[width=3.5cm]{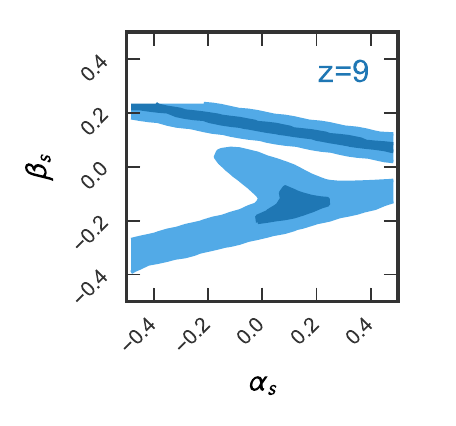}
\includegraphics[width=3.5cm]{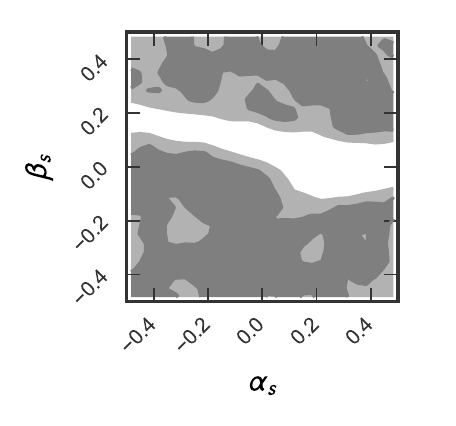}
\caption{\label{fig:mcmcabeachz} These panels show constraints on $\alpha_s$ and $\beta_s$ from the UV LFs at $z=6$, 7, 8, 9, and 10 from left to right. 
68 \% and 95\% allowed regions are depicted as dark and light colors, respectively.  These constraints are derived by performing the MCMC analysis only at each redshift. }
\end{figure*}

Fig.~\ref{fig:mcmcabeachz} shows running parameters constrained by using LFs for each redshift. These panels indicate that the constraints shown in Fig.~\ref{fig:nab} are dominated by those from the LFs at $z \leq 8$. For example, cases with
$(n_s, \alpha_{s}, \beta_s)=(0.97, 0.2, 0.0)$, and $(n_s, \alpha_{s},\beta_s) =(0.97, 0.0, -0.1)$ are ruled out because these models enhance and reduce the LFs overall as shown in Fig.~\ref{fig:nabLF}. This is due to the fact that the running parameters change the overall PPS amplitude around the relevant scale ($1 <k< 10 \rm Mpc^{-1}$), and hence the enhancement or the suppression of the LFs is independent of the galaxy magnitude $M_{\rm UV}$.

%%%%%%%%%%%%%%%%%%%%%%%%%%%%%%%%%
\subsection{\label{sec:const_param_II} Constraints on PPS: Parametrization II \\(step function-like PPS)}

\begin{figure}[h]
\includegraphics[width=7.5cm]{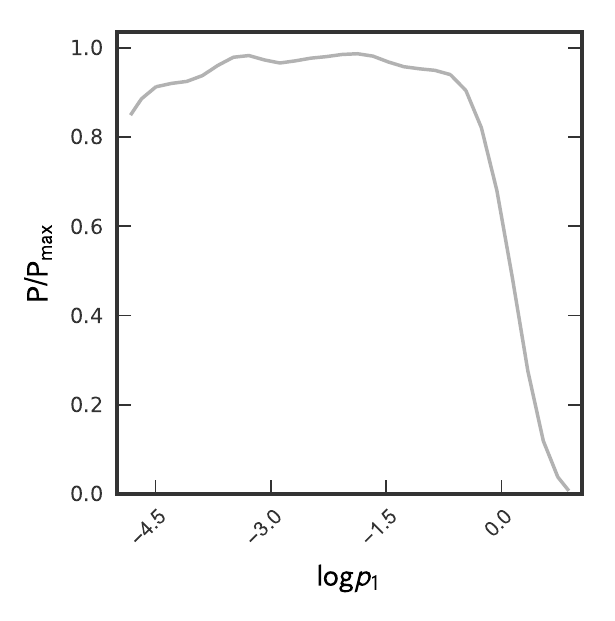}
\caption{\label{fig:mcmcp1} The posterior distribution of $p_1$ for Parametrization II with other parameters marginalized. This constraint is derived using all the UV LFs. }
\end{figure}

Next we show our results for the case of Parametrization II. The PPS is described as a step function-like form at $k>10 \rm Mpc^{-1}$. As in Parametrization I, we use the LFs at all redshift and the prior on the $n_s$  to constrain 22 free parameters. 
The constraint on the $p_1$ is shown in Fig.~\ref{fig:mcmcp1}. 
We find that $p_1$ is consistent with 0 with the 95\% upper limit of
\begin{eqnarray}
\log p_1 < 0.10. %%95% 
%0.290 %99%.
\end{eqnarray}
The constraint is stronger than previous constraints on the PPS (e.g. Fig.~6 in \cite{Bringmann:2011ut}). The UV LFs cannot place the lower limit on $p_1$ because ${\cal P_\zeta }(k)$ at large scales dominates the mass variance and hence, even if small scale amplitude gets significantly suppressed, the LFs do not decrease even for extremely low values  $p_1$.

\begin{figure*}[!t]
\includegraphics[width=3.5cm]{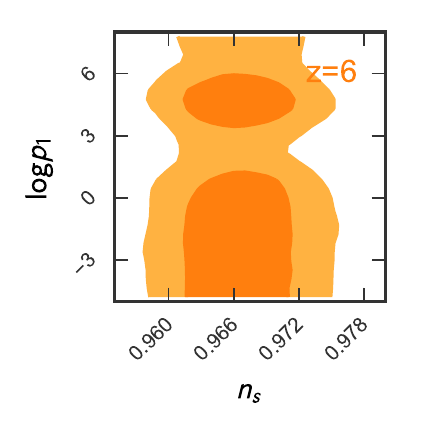}
\includegraphics[width=3.5cm]{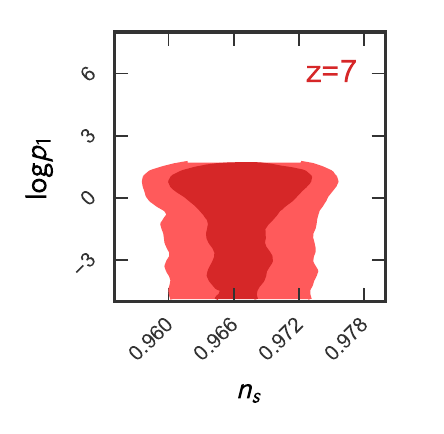}
\includegraphics[width=3.5cm]{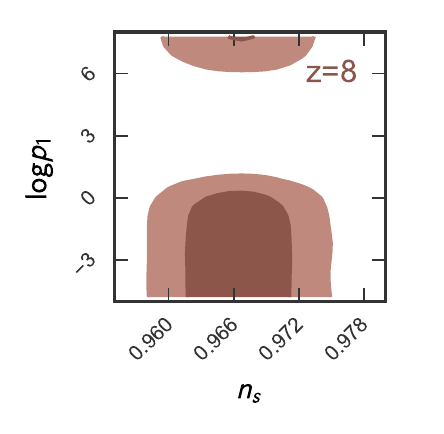}
\includegraphics[width=3.5cm]{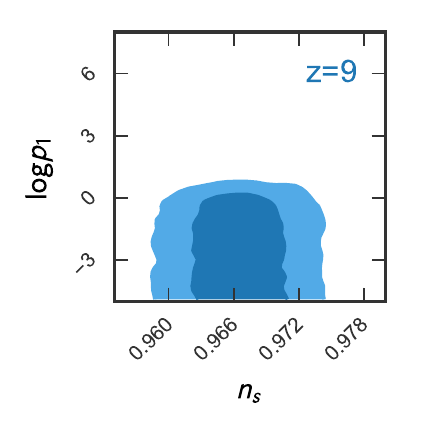}
\includegraphics[width=3.5cm]{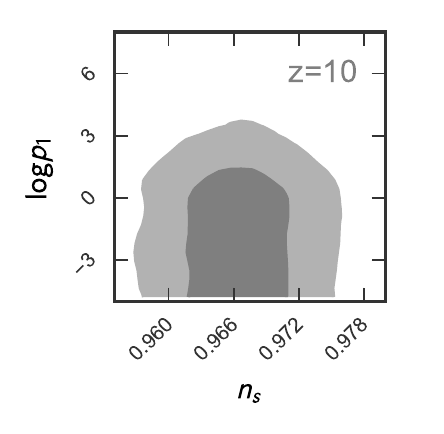}
\caption{ \label{fig:mcmcp1eachz} 
Same as Fig.~\ref{fig:mcmcabeachz}, but for constraints on $n_s$ and $p_1$ for Parametrization II.}
\end{figure*}

In the same way as in Parametrization I, in order to clarify LFs from which redshift dominate the constraints, we show constraints on $p_1$ and $n_s$ from the LFs at each redshift in Fig.~\ref{fig:mcmcp1eachz}. The panels show that the constraints are dominated by those from the UV LF at $z=9$. Note that extremely large $p_1$ is allowed at $z=6$ and $z=8$ if $f_{\rm star}$ is sufficiently small. In this case, other astrophysical parameters take values of $\log f_{*,10}\approx -2.5$ and $\alpha_{\rm star}=0.0$. However, such models are rejected because the halo mass function becomes too high even at $z\geq 9$. 

%%%%%%%%%%%%%%%%%%%%%%%%%%%%%%%%%
\subsection{\label{sec:const_param_III} Constraints on PPS: Parametrization III (binned PPS)}

\begin{figure*}[!t]
\includegraphics[width=5.5cm]{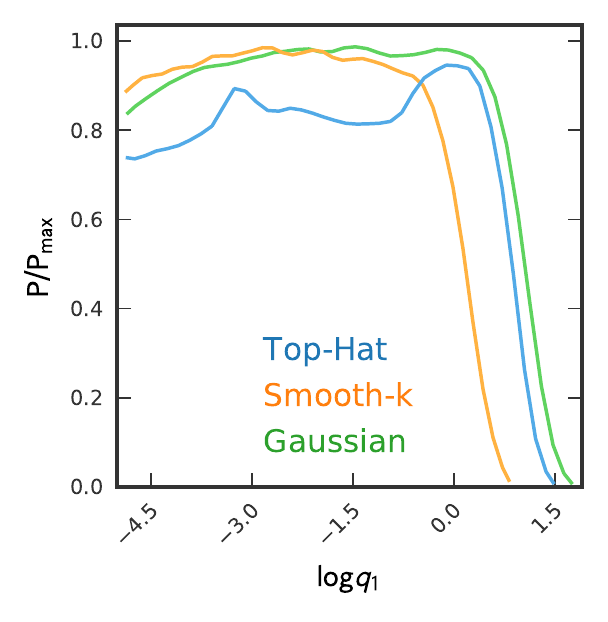}
\includegraphics[width=5.5cm]{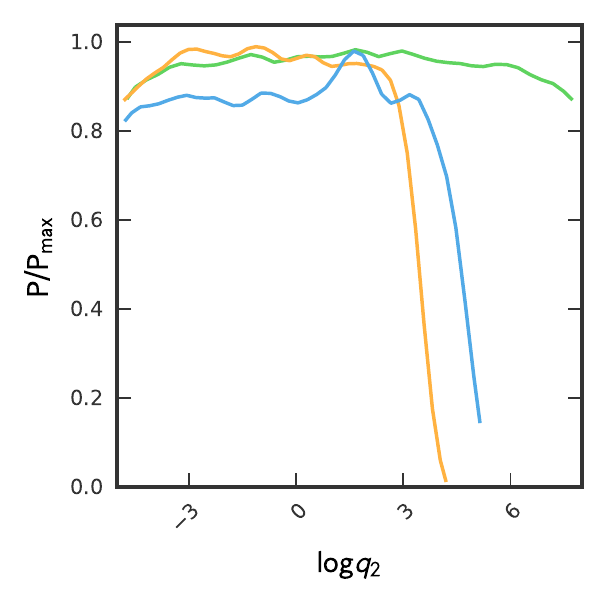}
\includegraphics[width=5.5cm]{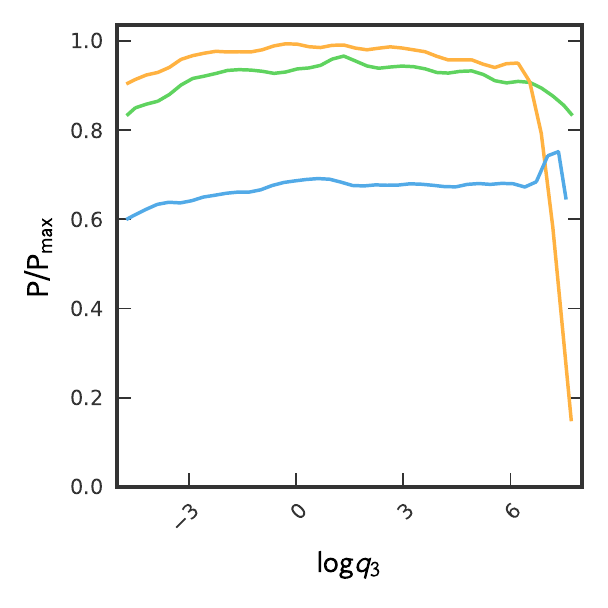}
\caption{\label{fig:param3upper} One-dimensional posterior distributions for $\log q_1$, $\log q_2$, and $q_3$ for Parametrization III. Blue, orange, and green lines are the result with top-hat, smooth-$k$, and Gaussian filters,  respectively.}
\end{figure*}

For the case of Parametrization III, the PPS is described by a binned step functions-like form. The one-dimensional posterior distributions for $q_1$, $q_2$, and $q_3$, which characterize the amplitude of each bin, are shown in Fig.~\ref{fig:param3upper}. 
For the top-hat filter, 95 \% upper bounds are obtained as
\begin{eqnarray}
\log q_1<0.2, \notag \\  [5pt]
\log q_2<3.2,  \notag \\  [5pt]
\log q_3<6.9.
\end{eqnarray}
We note that our constraint on PPS at $k<100 \rm Mpc^{-1}$ is stronger than those obtained in previous work. Constraints on astrophysical parameters are listed in Table.~\ref{tab:mcmcz} and those on PPS parameters are given in  Table~\ref{tab:mcmcpps}. The full corner plot is shown in Appendix~\ref{sec:Afr} (Fig.~\ref{fig:mcmcMC2}).

Examples of LFs are shown in Fig.~\ref{fig:LFall}, where we compare the LFs with $\log q_{1,2,3}=0$, $\log q_{1,2,3}=-5$ and a model with 95 \%  upper limit  of $q_{1,2,3}$. The figure shows that the small scale PPS controls the faint end of LFs. Thus, future improved constraints on the faint end of LFs have a potential to tighten constraints on PPS at small scales.

Interestingly, the best-fit value of $\alpha_{\rm star}$ depends on redshift, which hints the redshift evolution of these parameters. Therefore the evolution of these astrophysics parameters should be taken into account in future parameter estimations including those via 21cm line. However, the understanding the evolution of such astrophysical parameters is beyond the scope of this work. 

% \begin{figure*}[!t]
% \includegraphics[width=15.5cm]{figure/test_z7z8.pdf}
% \caption{\label{fig:mcmcz} Result of MCMC at mutiple redshift. Iteration 15000, 64 walkers.  We use python package sampler emcee with sampling moves are default. The LFs at z=10 does not constraint the parameters at all. (SY : I found useage of Oesch 2015 was wrong in the MCMC. I'll correct it soon. I'll add the result with $z$=6 later.) In this result we assume smooth-k filter. }
% \end{figure*}

\begin{table*}[!t]
\caption{\label{tab:mcmcz}
Summary of constraints on astrophysical parameters obtained from MCMC analysis using UV LFs from all redshifts and the prior on the $n_s$ from Planck for Parametrization III. Here we assume the smooth-$k$ filter. }
\begin{ruledtabular}
\begin{tabular}{cccccc}
 &$z=6$ &$z=7$ &$z=8$ &$z=9$ & $z=10$\\
%\mbox{Three}&\mbox{Four}&\mbox{Five}\\
\hline
$\alpha_{\rm star}$ &
$ 0.301_{-0.154}^{0.188} $ & 
$ 0.342_{-0.080}^{0.097} $ & 
$ 0.834_{-0.167}^{0.115} $ & 
$ 0.788_{-0.221}^{0.149} $ &
$ 0.302_{-0.499}^{0.459} $ \\ [5pt]
$\log f_{*,10}$ & 
$ -1.204_{-0.419}^{0.238} $ &
$ -1.156_{-0.354}^{0.200} $ &
$ -1.531_{-0.410}^{0.226} $ &
$ -1.138_{-0.435}^{0.247} $ &
$ -2.129_{-0.567}^{0.545} $ \\  [5pt]
$t_{\rm star}$ &
$ 0.522_{-0.320}^{0.317} $ &
$ 0.547_{-0.305}^{0.300} $ &
$ 0.525_{-0.319}^{0.313} $ &
$ 0.512_{-0.323}^{0.321} $ &
$ 0.580_{-0.323}^{0.285} $\\  [5pt]
$\log M_{\rm turn}$ &
$ 8.965_{-0.642}^{0.605} $ &
$ 8.673_{-0.453}^{0.521} $ &
$ 9.203_{-0.787}^{0.589} $ &
$ 9.175_{-0.763}^{0.601} $ &
$ 9.022_{-0.680}^{0.663} $\\  [5pt]
%$\log p_1$ &  & $ -2.122_{-1.967}^{1.784} $ & $ -2.307_{-1.817}^{1.915} $ & $ -2.304_{-1.843}^{1.889} $ & $ 1.304_{-4.325}^{5.180} $\\
%$\log p_2$ &  & $ -0.775_{-2.915}^{3.024} $ & $ -0.547_{-3.008}^{2.948} $ & $ -0.721_{-2.907}^{2.758} $ & $ 0.936_{-4.032}^{4.468} $\\
%$\log p_3$ &  & $ 0.927_{-4.091}^{4.296} $ & $ 1.035_{-4.167}^{4.378} $ & $ 0.764_{-3.926}^{4.135} $ & $ 1.245_{-4.160}^{4.473} $\\
%$\log p_1^{up}$ &  & 0.775 & 0.830 & 0.949 &7.909\\
%$\log p_2^{up}$ &  & 3.792 & 4.043 & 3.811 &7.849\\
%$\log p_3^{up}$ &  & 7.198 & 7.584 & 7.194 &7.847\\
\end{tabular}
\end{ruledtabular}
\end{table*}

\begin{table*}[!t]
\caption{\label{tab:mcmcpps}
Summary of constraints on PPS parameters for Parametrization III obtained from the MCMC analysis using UV LFs from all redshifts. Here we present the results for each filter. Since the posterior distributions of $q_i$ is not Gaussian for all $i$, we only show the 95\% upper limit $q_i^{\rm up}$. Note that we do not obtain an upper bound for $q_2$ and $q_3$ when top-hat and Gaussian filters are adopted.  }
\begin{ruledtabular}
\begin{tabular}{cccccc}
 &Smooth-$k$ &Top-hat &Gaussian \\
%\mbox{Three}&\mbox{Four}&\mbox{Five}\\
\hline

% $\log q_1$ & 
% $ -2.406_{-1.727}^{1.733} $ &
% $ -1.945_{-1.994}^{1.968} $ &
% $ -1.941_{-2.011}^{2.030} $ \\
% $\log q_2$ &
% $ -0.789_{-2.799}^{2.869} $ &
% $ -0.095_{-3.301}^{3.208} $ &
% $ 1.494_{-4.345}^{4.362} $ \\
% $\log q_3$ &
% $ 1.183_{-4.132}^{4.129} $ &
% $ 1.393_{-4.267}^{4.256} $ &
% $ 1.522_{-4.321}^{4.310} $  \\

%% 99%
% $\log q_1^{up}$ & 0.344 & 0.946 & 1.166 \\
% $\log q_2^{up}$ & 3.544 & 4.833 & - \\
% $\log q_3^{up}$ & 7.311 & - & - \\

%% 95%
$\log q_1^{up}$ & 0.15 & 0.57 & 0.81 \\  [5pt]
$\log q_2^{up}$ & 3.18 & 4.42 & - \\  [5pt]
$\log q_3^{up}$ & 6.83 & - & - \\

\end{tabular}
\end{ruledtabular}
\end{table*}

The constraints depend on the choice of the filter. In Fig.~\ref{fig:param3upper}, we compare the constraints for various filters. The result indicates that the constraints are tighter when the smooth-$k$ filter is assumed. This is due to the fact that the smooth-$k$ filter allows small scale PPS to contribute to the mass variance as shown in Fig.~\ref{fig:MUV2Mhalo2}. 
However, even in the case of the top-hat filter, the constraint on the PPS at $k<100~ \rm Mpc^{-1}$ is tighter than in previous work \cite{Bringmann:2011ut} (in our case ${\cal P}_\zeta (10 <k< 100 ~{\rm Mpc}^{-1})  < {\cal O}(10^{-8}) $).  Note that the top-hat filter can show artificial features in the LFs for models with extremely large $q_i$ ($i$=1, 2, 3) as shown in \cite{Yoshiura:2019zxq}.
In contrast, since the Gaussian filter erases the small scale power completely, the constraint is weak. The UV LFs cannot constrain $q_2$ and $q_3$, but the $q_1$ should be less than ${\cal O}(10)$, i.e., ${\cal P}_\zeta (10 <k< 100 ~{\rm Mpc}^{-1})  < {\cal O}(10^{-8}) $.
We again caution that the appropriate filter should ultimately be chosen by comparing the mass function with $N$-body simulation results with such an enhanced PPS model.

The constrained astrophysical parameter for each filter are consistent within 1$\sigma$ error while the best fit value of $\alpha_{\rm star}$ and $f_{*,10}$ slightly depend also on the choice of the filter (see Fig.~\ref{fig:mcmcfilter} in Appendix~B).
We will discuss the effect of the cosmological assumption on the astrophysical parameters in later section. % because the astrophysical parameter can degenerate with $q_i$ as discussed. %This would affect the future parameter constraining using reionization model via 21cm line. 

%%%%%%%%%%%%%%%%%%%%%%%%%%%%%%%%%
\subsection{\label{sec:const_param_IV} Constraints on PPS: Parametrization IV (scale dependent step function-like PPS)}  

\begin{figure}[h]
\includegraphics[width=7.5cm]{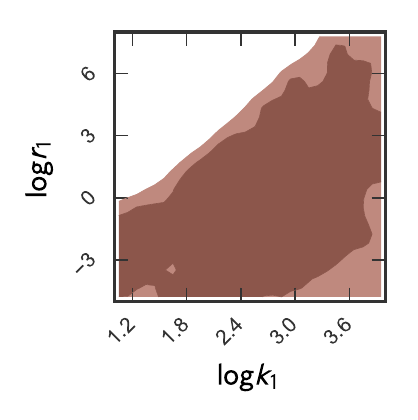}
\caption{\label{fig:p1k1} Constraints on $r_1$ and $k_1$ for Parametrization IV from the MCMC analysis using UV LFs from all redshifts.
The smooth-$k$ filter is adopted in this figure.
68 \% and 95\% allowed regions are depicted as dark and light colors, respectively. }
\end{figure}

We show the result for Parametrization IV in Fig.~\ref{fig:p1k1}. This parametrization describes a step function-like form with an arbitrary transition scale. In this case, the upper limit on the PPS can be estimated at various scales. In Fig.~\ref{fig:p1k1}, the corner plot of $r_1$ and $k_1$ shows that the larger $k_1$ is, the larger fluctuation is allowed. This figure also indicates that upper limits on the PPS are placed for a wide range of scales. 
\begin{figure*}[!t]
\includegraphics[width=20.0cm]{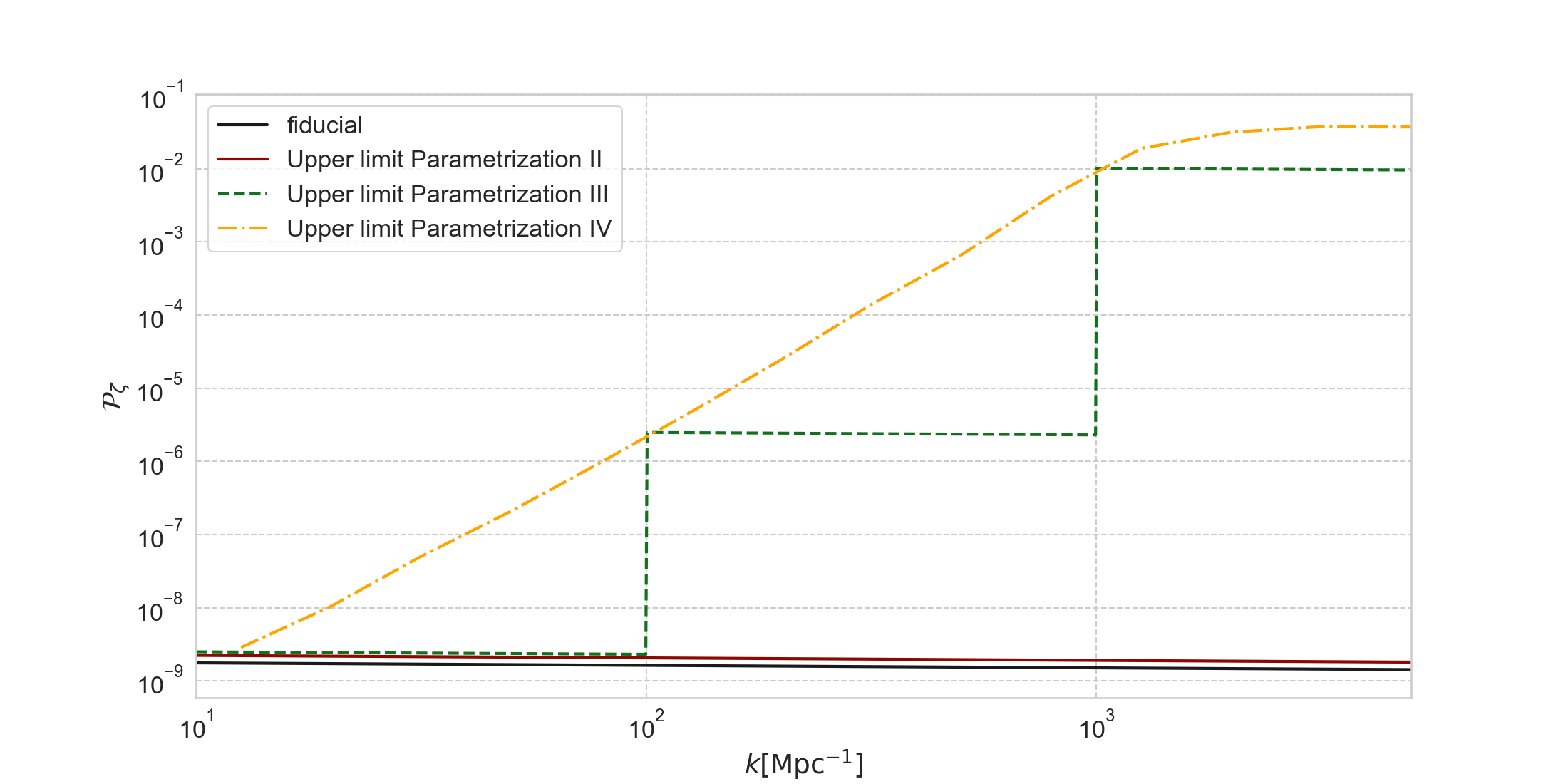}
\caption{\label{fig:1d} Upper limits on PPS for various parmetrizations for the smooth-$k$ filter. Upper solid, dashed, and dot-dashed lines show the 95\% upper bounds for the parametrization II, III, and IV, respectively. We plot PPS with $\log q_{i=1,2,3}=0$ as fiducial.}
\end{figure*}

%%%%%%%%%%%%%%%%%%%%%%%%%%%%%%%%%
\subsection{\label{sec:const_param_V} Constraints on PPS: Parametrization V (power-law PPS)}  

In Parametrization V, only $n_s$ is allowed to vary. The $n_s$ is severely constrained by the Planck. Thus, we can obtain best fit astrophysical parameters without PPS uncertainty.
The constraints on the astrophysical parameters are fairly consistent with the values shown in \cite{Park2019MNRAS.484..933P}.  The full result is shown in Fig.~\ref{fig:mcmcMC5} of  Appendix~\ref{sec:Afr}. %The difference of parameters at different redshift might indicate the evolution of these parameters, but further investigation is out of scope of this paper. 

%\bigskip\bigskip\bigskip
%%%%%%%%%%%%%%%%%%%%%%%%%%%%%%%%%
\subsection{Summary of constraints on PPS}

In Fig.~\ref{fig:1d}, we summarize the 95\% upper bounds obtained in Parametrization II, III, and IV for the smooth-$k$ filter. For Parametrization IV, we assort MCMC samples into several $k$-bins and calculate 95\% upper limits from the samples in each $k$-bin, which almost corresponds to the upper bound in Fig.~\ref{fig:p1k1}. 
We find that our results for Parametrization III and IV are reasonably well consistent with each other. It is also found that the analysis for Parametrization II severely overestimate the upper limits on PPS at very small scales.

%%%%%%%%%%%%%%%%%%%%%%%%%%%%%%%%%%%%%%%%%%%%%%%%%%%%%%%%%%%%%%%%%%
%%%%%%%%%%%%%%%%%%%%%%%%%%%%%%%%%%%%%%%%%%%%%%%%%%%%%%%%%%%%%%%%%%
\section{Discussions \label{sec:discussions}}

%%%%%%%%%%%%%%%%%%%%%%%%%%%%%%%%%
\subsection{Interpretation of the constraints from UV Magnitude}

\begin{figure}[h]
\includegraphics[width=9.5cm]{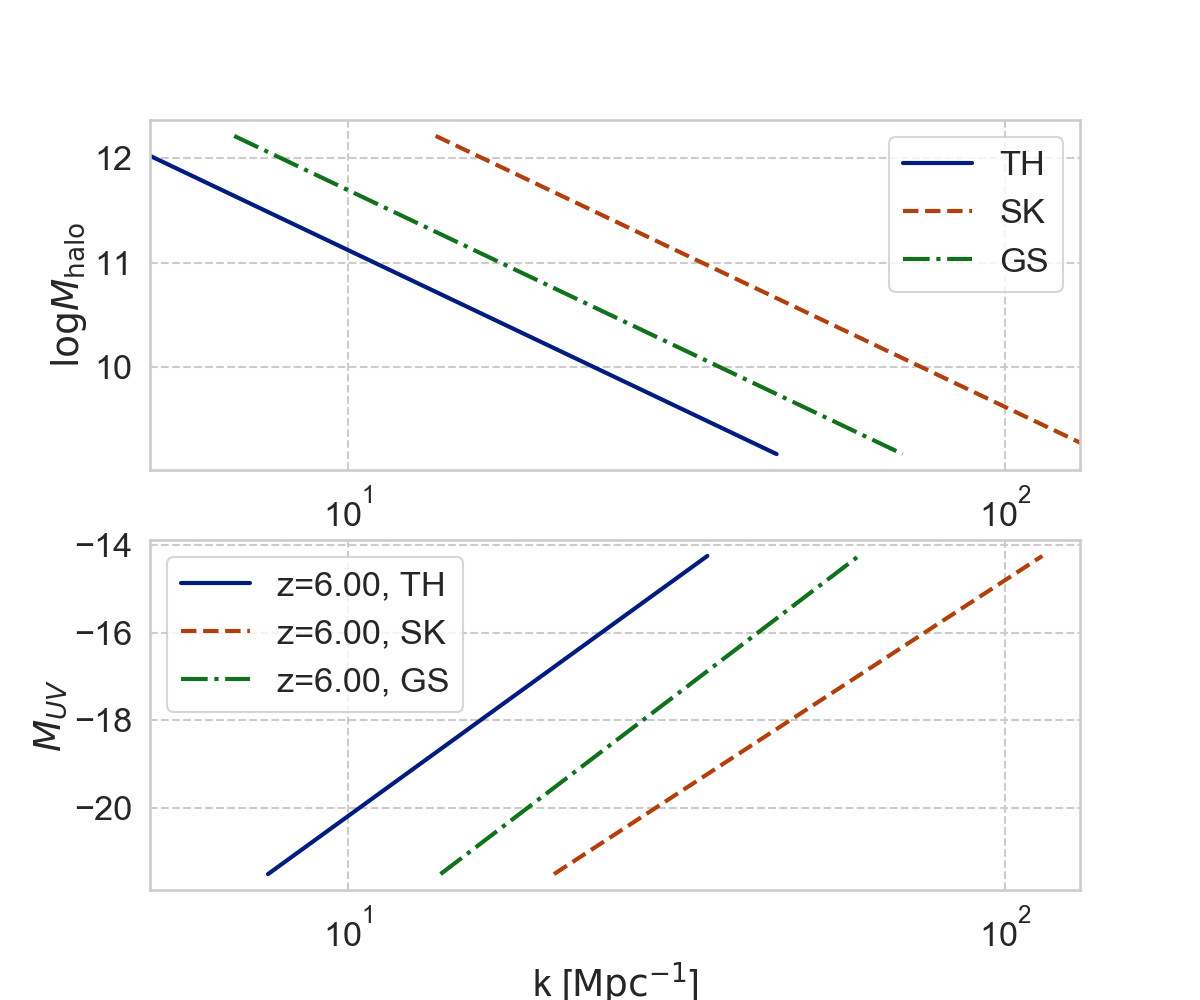}
\caption{\label{fig:MUV2Mhalo1} {\it Upper}: Relation between halo mass $M_{\rm halo}$ and the wavenumber $k$ for each filter at $z=6$. {\it Lower}: Relation between galaxy UV magnitude $M_{\rm UV}$ and $k$ for each filter at $z=6$ assuming the best-fit astrophysical parameters for Parametrization III. }
\end{figure}

In order to interpret the constraints, it is useful to see  which scale $k$ is responsible for a halo with $M_{\rm halo}$ and UV magnitude $M_{\rm UV}$. 
$M_{\rm halo}$ and $k$ are related via Eqs.~\eqref{eq:mass_scale}, \eqref{eq:mass_scale_gaussian}, and \eqref{eq:mass_scale_smoothk} for top-hat, Gaussian, and smooth-$k$ filters, respectively.
$M_{\rm UV}$ is given for a fixed $k$ from the formulas given in Section~\ref{sec:mass_function}.   In the upper panel of Fig.~\ref{fig:MUV2Mhalo1}, we show the $M_{\rm halo}$-$k$ relation, which indicates that the smooth-$k$ filter allows us to explore the PPS on smaller scales than in other filters. We also show the $M_{\rm UV}$-$k$ relation in the bottom panel of Fig.~\ref{fig:MUV2Mhalo1} with the best-fit astrophysical parameters for each filter derived in Parametrization III. This panel indicates that observations of faint-end galaxies down to $M_{\rm UV} \approx -14$ can probe the scales of $k\approx 100 \rm Mpc^{-1}$ at $z=6$. This is why the UV LFs can constrain $q_1$ and $q_2$ for  Parametrization III.

Not only the mass scale but also the shape of the filter as a function of $k$ is crucial for understanding the constraints. Fig.~\ref{fig:MUV2Mhalo2} indicates that the smooth-$k$ filter has a strong response to small scales, and this is why the constraint with the smooth-$k$ filter is tighter than in other ones. The Gaussian filter, on the other hand, does not pick up fluctuations on small scales. Thus, the PPS at $k>20 \, \rm Mpc^{-1}$ cannot be constrained with the Gaussian filter. %Note that the shape of smooth k filter depends on $\beta$. When $\beta=2$, the response to small scales becomes weaker.

\begin{figure}[h]
\includegraphics[width=9.5cm]{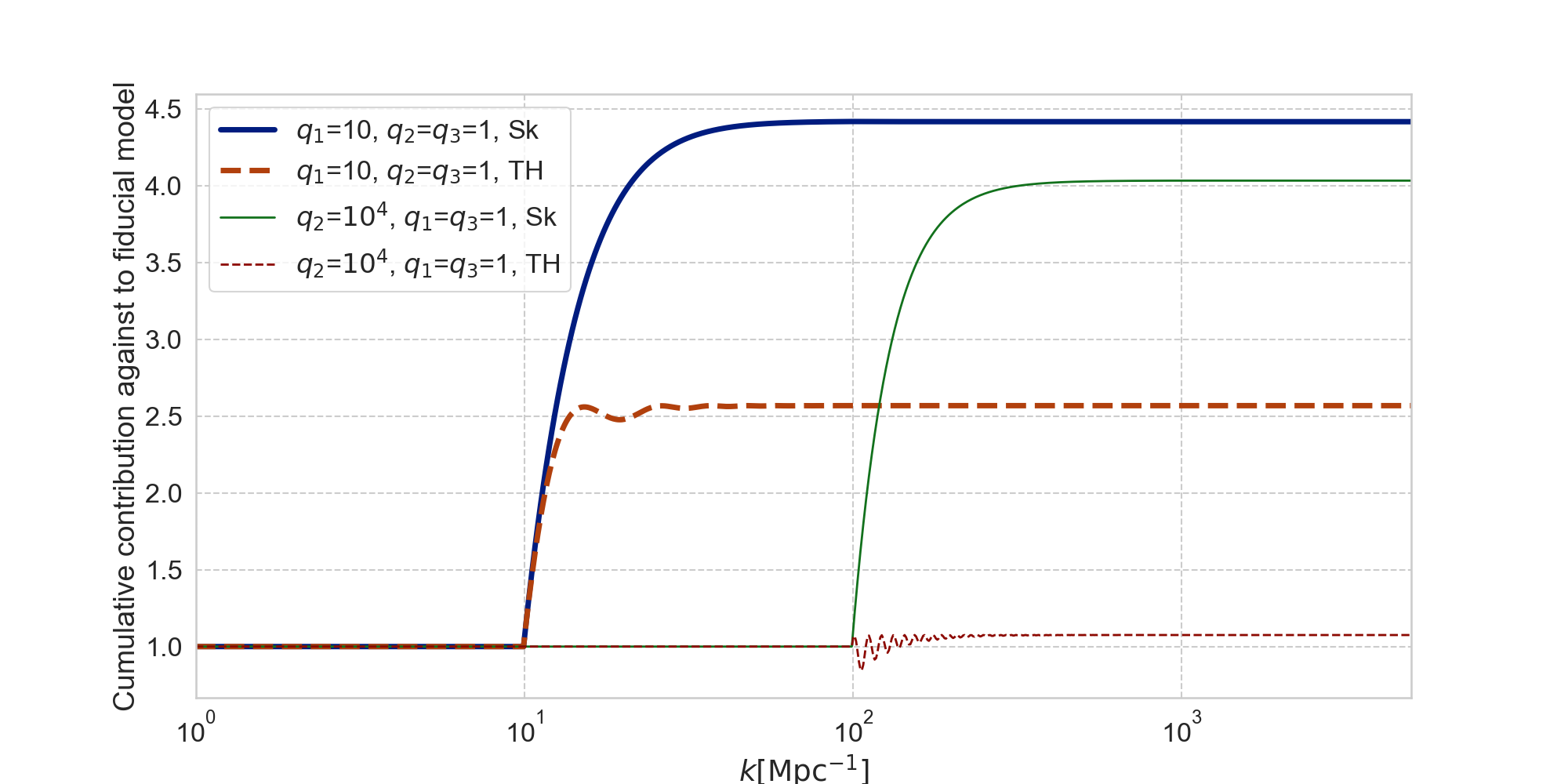}
\caption{\label{fig:MUV2Mhalo3} Cumulative value of Eq.~\eqref{eq:dlnsigmadlnm} for enhanced PPS models. For comparison, we plot the ratio against to $p_1=p_2=p_3=1$ model. Dashed lines show the result with the top-hat filter, and solid lines are the result with the smooth-$k$ filter.}
\end{figure}

The integral of Eq.~\eqref{eq:dlnsigmadlnm} controls the contribution to the mass function from fluctuations at small scales. Thus, a cumulative contribution of the integral up to $k$ mode is useful to estimate the contribution from particular mass scales. Fig.~\ref{fig:MUV2Mhalo3} shows the ratio of cumulative contribution for PPS with Parametrization~III against to a fiducial model with $q_1=q_2=q_3=1$. In the figure, we show the cumulative contribution for a galaxy with $M_{\rm UV}=-15$ which corresponds to $M=10^{9.6} M_\odot$.  This figure shows that the smooth-$k$ filter picks up more power from the enhanced PPS on small scales.

For a model with $q_1=10$ ($q_2=10^4$), the total contribution from the smooth-$k$ filter in the range of $10~{\rm Mpc}^{-1} <k<100~{\rm Mpc}^{-1}$ ($100~{\rm Mpc}^{-1} <k<1000~{\rm Mpc}^{-1}$) is 4.5 (4.0) times larger than the fiducial model. 
On the other hand, in the case of the top-hat filter, the enhancement of the total contribution is smaller than in the smooth-$k$ filter. This is why the top-hat filter gives weaker upper limits on $q_i$ ($i$=1, 2, 3) compared to those for the smooth-$k$filter.

%%%%%%%%%%%%%%%%%%%%%%%%%%%%%%%%%
\subsection{\label{sec:SHMR} Impact of stellar to halo mass ratio}

We have also included the data of the stellar to halo mass ratio (SHMR) \cite{2016ApJ...821..123H}  in our MCMC analysis to constrain the parameters. Although the SHMR relation is constrained only at the bright end of the LFs ($M_{\rm UV}\approx -20$), it still helps improve constraints on PPS by constraining the astrophysical parameters. A caveat is that constraints on SHMR from the clustering analysis in principle depend on PPS. However, the SHMR is mainly constrained from the so-called 2-halo term of the galaxy auto correlation function at $\theta > 20''$, which corresponds to $k<8  \,\rm Mpc^{-1}$ at $z=6$. This $k$ range is smaller than that of our interest, and therefore for simplicity we assume that the constraints on the SHMR in \cite{2016ApJ...821..123H}  does not depend on PPS and thus do not change throughout the MCMC analysis. 

\begin{figure*}[!t]
\includegraphics[width=18cm]{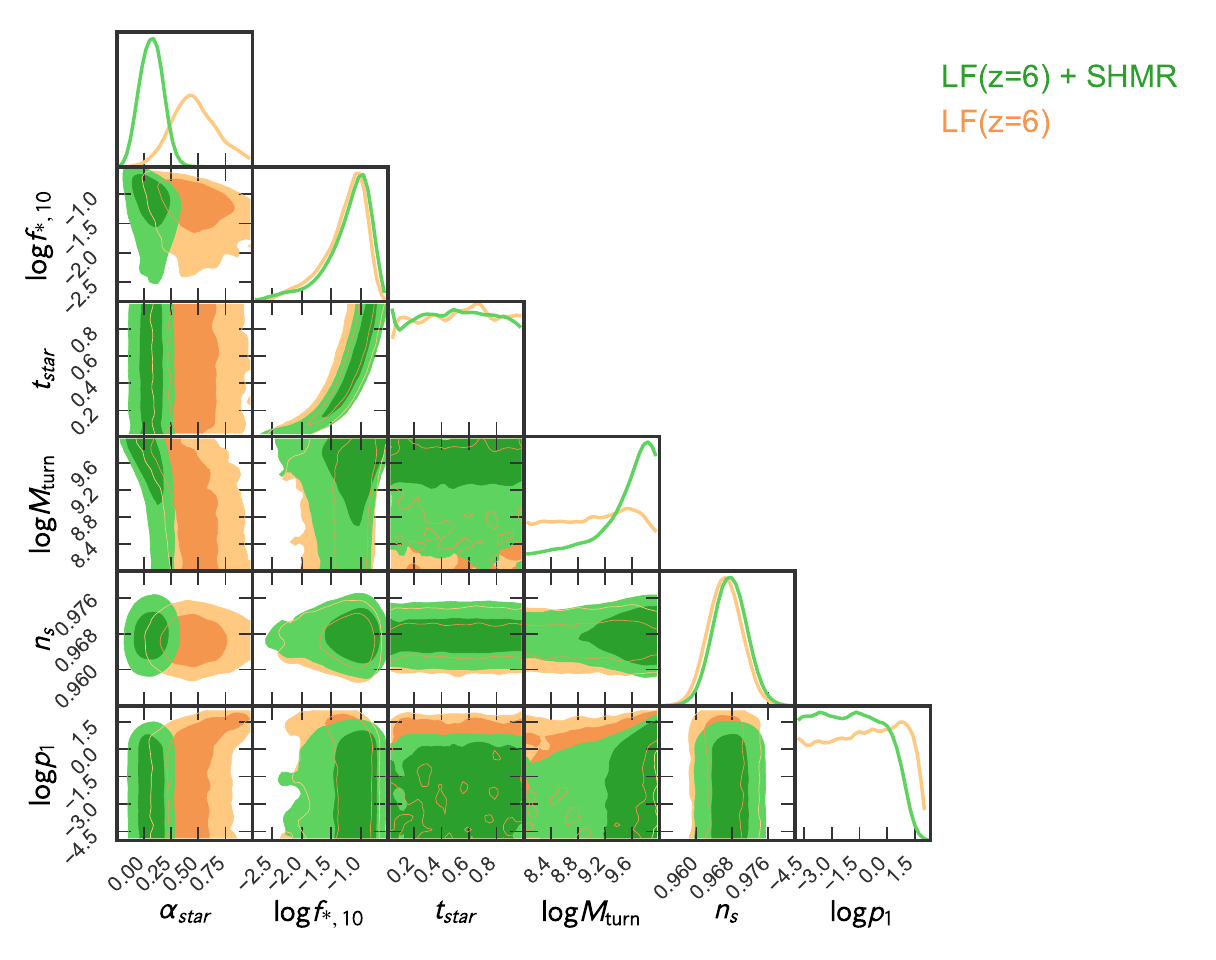}
\caption{\label{fig:mcmcHY} Result for the MCMC analysis for Parametrization II using only the UV LF at $z=6$ (orange) and both the UV LF and SHMR at $z=6$ (green).}
\end{figure*}

\begin{figure*}[!t]
\includegraphics[width=15.5cm]{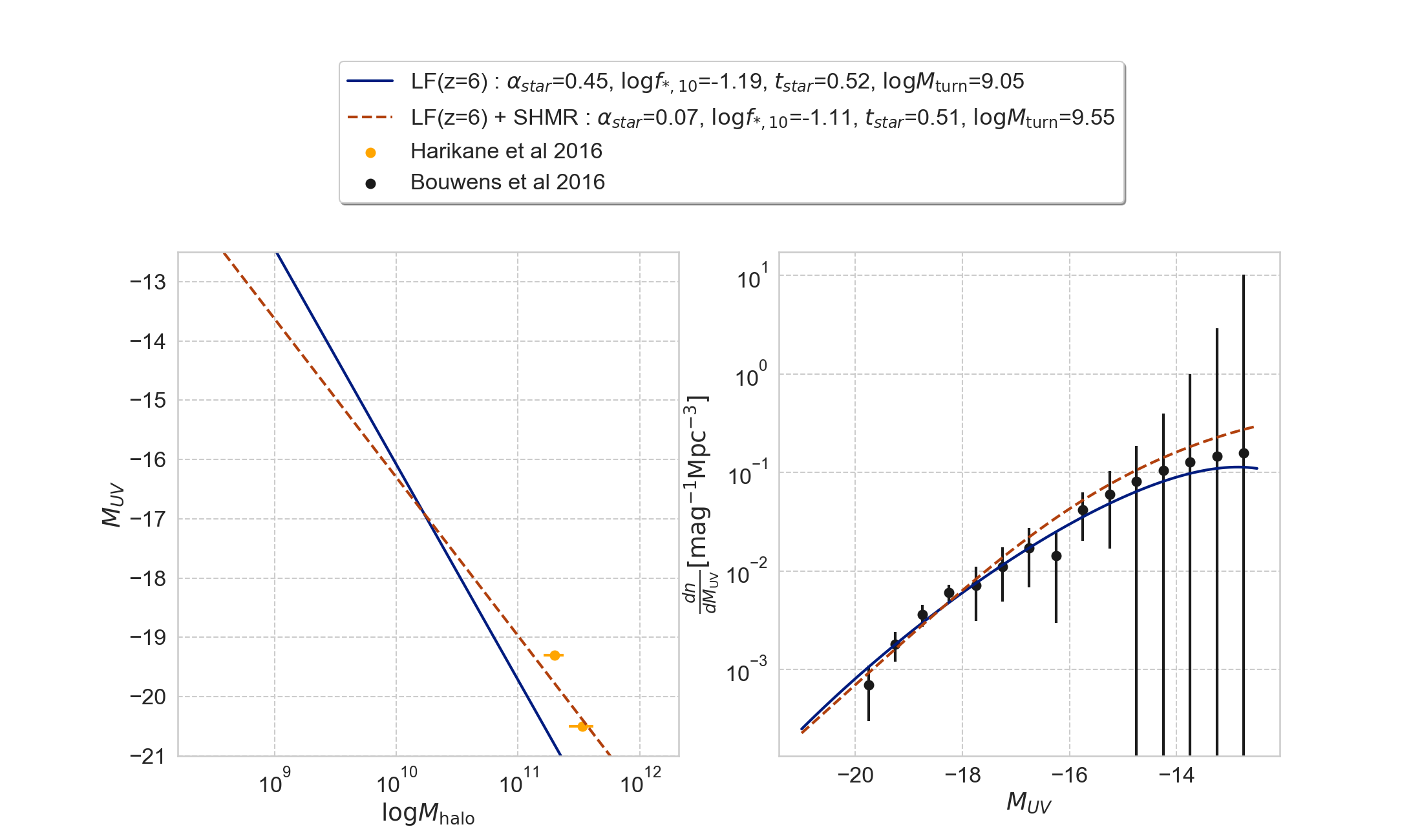}
\caption{\label{fig:LFHY} {\it Left}: Relation between $M_{\rm UV}$ and $M_{\rm halo}$ at $z=6$. The observational constraints from \cite{2016ApJ...821..123H} are compared with best-fit models  from the UV LF at $z=6$ only ({\it solid}) and from both UV LF and SHMR ({\it dashed}). See Tabel~\ref{tab:astroHY} for the best-fit parameters. {\it Right}: Comparison of the observed UV LF at $z=6$ from \cite{Bouwens2017ApJ...843..129B} with best-fitting models also shown in the left panel. Here we fix $p_1=1$.}
\end{figure*}

To check this point explicitly, in Fig.~\ref{fig:mcmcHY} we show results of MCMC using the UV LF at $z=6$ and/or the SHMR with the top-hat filter for PPS with $p_1$ (Parametrization II). In Table~\ref{tab:astroHY}, we list constraints on the astrophysical parameters
as well as on the PPS parameter. We find that the best-fit value of $\alpha_{\rm star}$ varies somewhat by adding the SHMR data. In order to understand this result, we plot the SHMR and the UV LF at $z=6$ using the best-fit parameters in Fig.~\ref{fig:LFHY}. The left panel indicates that, in order to reproduce the observed SHMR, star formation rate (SFR) in halo of $M_{h} \approx 10^{11.5} M_{\odot}$ has to be suppressed. Here, the SHMR is described as $L_{\rm UV}\propto f_{*,10}M^{\alpha_{\rm star}+1}_{h}$. However, $f_{*,10}$ is required to be $\approx$ 0.1 for reproducing the LF at $M_{\rm UV}\approx -17$. Thus, the slope $\alpha_{\rm star}$ becomes flatter to maintain the UV LF and to satisfy the constraints on the UV LF and the SHMR simultaneously. Since $\alpha_{\rm star}$ becomes flatter, the lower halo mass scales can be explored at faint-end of LFs. Thus, the constraints on the astrophysical parameters and $p_1$ become slightly tighter.

%The SHMR is described as $L_{\rm UV}\propto f_{*,10}M^{\alpha_{\rm star}+1}_{h}$.
%Using only SHMR for the MCMC, the best fit value shows $\alpha_{\rm star}=0.68$ and $f_{*,10}=-2.1$, but this model underestimates the LF as seen from the right panel of Fig.~\ref{fig:LFHY}. 
%Thus, the slope $\alpha_{\rm star}$ becomes flatter to enhance the UV LF at the faint end and satisfying the constraints on the SHMR and the UV LF simultaneously.

By adding the SHMR to the MCMC analysis, the constraints on $\alpha_{\rm star}$ becomes tighter. Thus, the SHMR is useful to remove the degeneracy between $\alpha_{\rm star}$ and the mass dependence of the escape fraction of ionising photons which has been found in \cite{Park2019MNRAS.484..933P}. %Thus, if $\alpha_{\rm star}$ is well constraint using the SHMR, the degeneracy can be solved. 
% and further constraint on the EoR parameters. the constraint becomes strong, observation on the stellar to halo mass ratio can further constraint on the EoR parameters. Especially, if $\alpha_{\rm star}$ is well constraint, the degeneracy between the mass dependence of the escape fraction can be solved (\cite{Park2019MNRAS.484..933P}).

Interestingly, the higher $M_{\rm turn}$ value is preferred to reduce the LFs at $M_{\rm UV}\approx -15$ when we use the UV LF and the SHMR in the MCMC as shown in Fig.~\ref{fig:mcmcHY}. This is because the flatter $\alpha_{\rm star}$ increases the UV LF too much at $M_{\rm UV}\approx -15$ since lower mass halos can be responsible for the UV LF. 

%Interestingly, the best fit value of $\alpha_{\rm star}$ changes clearly. The MCMC coner plot is shown in Fig.~\ref{fig:mcmcHY}. In order to understand this result, we plot the stellar to halo mass ratio and UV LF at $z$=6 using the best fit parameters in Fig.~\ref{fig:LFHY}. The left panel shows that the slope $\alpha_{\rm star}$ becomes steeper due to SHMR and lower halo mass scales can be explored. Thus, the constraints on PPS becomes stronger than that without SHMR. 

%Note that the constraint on the astro physical parameters using LF at $z=6,7,8,10$(Bouwens et al and Oesch et al) is comparamble to the result shown in Park et al. (SY : Not perfectly consistent. May be something wrong, or the data and the restriction are different. Note the correlation between astrophysical parameters are more or less consistent with that shown in \cite{Park2019MNRAS.484..933P}.) 

\begin{table}[h]
\caption{\label{tab:astroHY} Summary of parameters obtained by MCMC using the UV LF at $z=6$ and/or SHMR.}
\begin{ruledtabular}
\vspace{4mm}
\begin{tabular}{cccccc}
parameter & LF($z=6$) + SHMR & LF($z=6$) \\ [6pt] 
%\mbox{Three}&\mbox{Four}&\mbox{Five}\\
\hline  
$\alpha_{\rm star}$ &$ 0.069_{-0.104}^{0.099} $ & $ 0.454_{-0.189}^{0.230} $ \\ [6pt]
$\log f_{*,10}$ &$ -1.113_{-0.460}^{0.245} $ & $ -1.192_{-0.456}^{0.249} $ \\ [6pt] 
$t_{\rm star}$ &$ 0.510_{-0.333}^{0.327} $ & $ 0.519_{-0.337}^{0.324} $ \\ [6pt] 
$\log M_{\rm turn}$ & $ 9.554_{-0.762}^{0.306} $& $ 9.053_{-0.706}^{0.605} $ \\ [6pt] 
$n_s$ &$ 0.968_{-0.004}^{0.003} $ & $ 0.966_{-0.003}^{0.004} $ \\ [6pt] 
%$\log p_{1}$ &$ -2.058_{-1.990}^{2.000} $ & $ -1.385_{-2.438}^{2.234} $& \\
$\log p_{1}^{up}$ & 0.71& 1.54\\ [6pt] 
\end{tabular}
\end{ruledtabular}
\end{table}

%%%%%%%%%%%%%%%%%%%%%%%%%%%%%%%%%
\subsection{Effects of cosmological assumptions on astrophysical parameters \label{sec:EOAA}}

Since we allow the form of PPS to vary especially on small scales,  constraints on astrophysical parameters may be affected due to the degeneracy with the PPS parameters. Here we discuss how the astrophysical parameter estimation depends on the assumption on cosmology.

\begin{figure*}[!t]
\includegraphics[width=6.5cm]{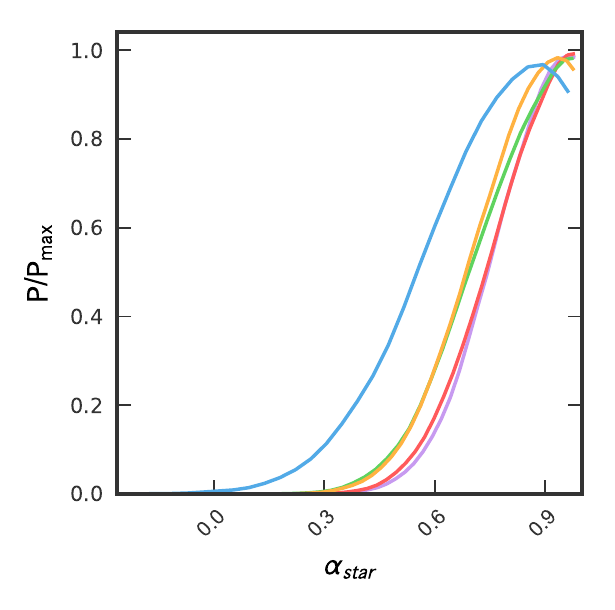}
\includegraphics[width=6.5cm]{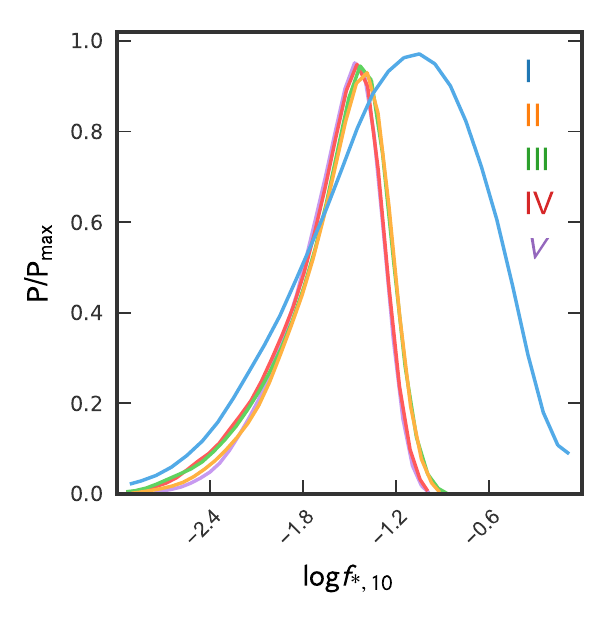}
\caption{\label{fig:mcmcabdegene1d} {Comparison of one-dimensional posterior distributions of $\alpha_{\rm star}$ ({\it left}) and $f_{*,10}$ ({\it right}) at $z=8$ with different parametrizations of PPS. We show results of the MCMC analysis using the UV LFs from all redshifts assuming the smooth-$k$ filter.} }
\end{figure*}

In Fig.~\ref{fig:mcmcabdegene1d}, we show the one-dimensional posterior distributions of $\alpha_{\rm star}$ and $f_{*,10}$ which are obtained using UV LFs from all redshifts to compare the results for all the parametrizations with the smooth-$k$ filter.

\begin{figure}[!t]
\includegraphics[width=7.5cm]{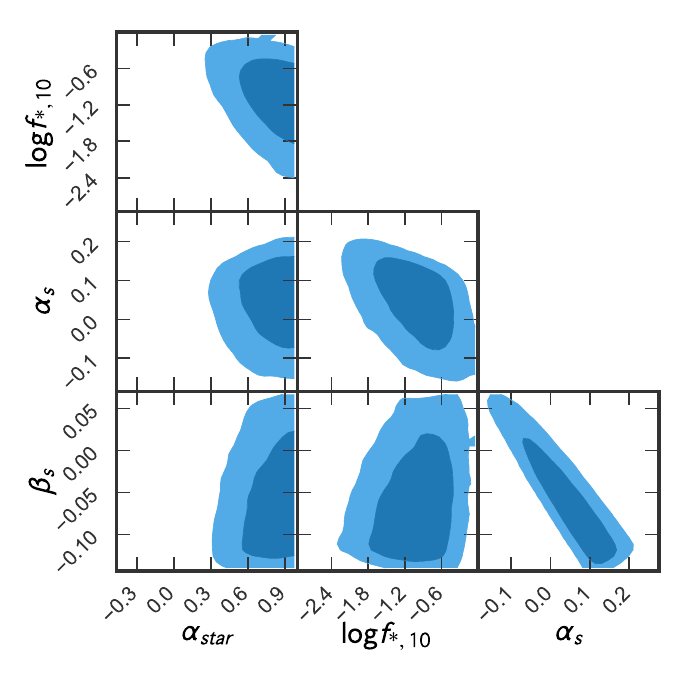}
\caption{\label{fig:mcmcabdegene} {Corner plot for Parametrization I. We show constraintns on the running parameters and astrophysical parameters $\alpha_{\rm star}$ and $f_{*,10}$) at $z=8$.}}
\end{figure}

For Parametrization I, $\alpha_{\rm star}$ and $f_{*,10}$ show weak correlations with $\alpha_s$ and $\beta_s$ as shown in Fig.~\ref{fig:mcmcabdegene}. Due to the degeneracy, the constraints on the astrophysical parameters are slightly loosen compared to the result of Parametrization V. 

\begin{figure}[!t]
\includegraphics[width=7.5cm]{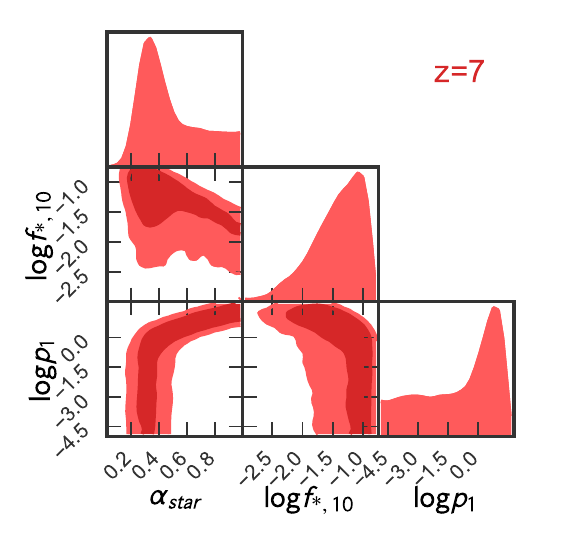}
\caption{\label{fig:mcmcz7} {Corner plot obtained by the MCMC analysis using UV LF at $z=7$ for Parametrization II.  }}
\end{figure}

On the other hand, the results of the other parametrizations of the PPS do not show any deviations from that of Parametrization V,  which has the standard simple power-law form when we include the data from all redshifts. However, when we perform MCMC using LFs only at $z=7$, a clear degeneracy is found between $p_1$, $f_{*,10}$, and $\alpha_{\rm star}$ as shown in Fig.~\ref{fig:mcmcz7}. Since large $p_1$ increases the halo mass function at small mass scales, low $f_{\rm star}$  with positive $\alpha_{\rm star}$ and lower  $f_{\rm *,10}$ 
can somewhat compensate the enhancement due to PPS. However this degeneracy does not affect Fig.~\ref{fig:mcmcabdegene1d} since the model with higher $p_1$ value is rejected by combining the UV LFs from other redshifts.

The redshift evolution of astrophysical parameter can also depend on the parametrization. For Parametrization II, III, and IV, the parameter evolution is same as that for Parametrization V. While the constraints become weak for Parametrization I (which is characterized by the runnings), the trend of the redshift evolution does not change. This can be seen in the full result of the MCMC analysis for each Parametrization shown in Appendix~\ref{sec:Afr}.

The best-fit values of astrophysical parameter $\alpha_{\rm star}$ and $f_{*,10}$ slightly depend on the choice of the filter as shown in Fig.~\ref{fig:mcmcfilter} , while the values for each filter are consistent with each other within 1 $\sigma$ error. For example, the Gaussian filter underestimate the halo mass function compared to other filters because the Gaussian filter erase the small scale contribution to the mass variance. Thus, in order to increase $f_{\rm star}$, the best fit value of $\alpha_{\rm star}$ becomes larger when the Gaussian filter is adopted. %This would affect the future parameter constraining using reionization model via 21cm line. 

%(SY : effect on EoR or ...? ) The assumption of PPS model affects the LFs at high-z as shown in \ref{fig:LFall}. In the context of EoR and Cosmic Dawn, the number of fainter galaxies are crucial to the evolution of thermal history and ionization history. 

%%%%%%%%%%%%%%%%%%%%%%%%%%%%%
\subsection{Model Uncertainties \label{sec:MUN}}

In this subsection, we discuss the uncertainties in the model of galaxy. First, dust attenuation should be taken into account carefully even for  high redshift LFs. For example, in \cite{2014MNRAS.440..731S, 2015MNRAS.451.2692C}, using numerical simulation results, they have shown that the LFs at $M_{\rm UV}<-19$ are decreased by the dust attenuation. However, the small scale PPS changes the LFs mainly at $M_{\rm UV}>-16$ as shown in Fig.~\ref{fig:LFall}. Thus, the dust effect on the PPS upper limits should be marginal, and thus we ignore the dust attenuation in our LF model.

We adapt a power law form for the stellar to halo mass ratio because the power law model is shown to be reasonable for faint galaxies as indicated in numerical simulations (e.g., \cite{2014MNRAS.440..731S,2017MNRAS.470.2791C,2018MNRAS.479..994R,2018MNRAS.478.1694M}). Although the model well reproduces the observed LFs, these assumptions on the functional forms might bias the constraints on the PPS if the true functional forms are different from those assumed forms.
Possible another choice of the functional form is a double power law form adopted in \cite{2018ApJ...868...92T}. However, they found the break mass of $M_{\rm halo}\approx 10^{11} M_\odot$, below which a single power law form is well consistent with radiation hydrodynamics numerical simulations. %in the range of low halo mass ($M_{\rm halo}<10^{11}$). 
Since we use relatively faint galaxies ($M_{\rm UV}>-20$), the assumption of the single power law is consistent with \cite{2018ApJ...868...92T}. The double power law model may be more appropriate when we use LFs of bright galaxies in the MCMC. We also note that, in Fig.14 of \cite{2018ApJ...868...92T}, numerical simulation results are not consistent with each other and the result shows that the stellar to halo mass ratio has large variance at small mass scales. Thus, as alternative form, for example, a stochastic implementation of the stellar mass--halo mass relation may be more suitable to predict the UV LF at the faint end. While it would be interesting to compare the results using other forms to describe the LFs more flexibly, we leave it for future work.   

Furthermore, modelling of astrophysical effects should involve various uncertanities. For example, internal and external UV backgrounds significantly suppress the star formation in low mass halos corresponding to the very faint end of current observations (e.g., $M_{\rm halo}<10^9 M_\odot$, \cite{2013MNRAS.428..154H}). In addition to the UV feedback, the relative velocity of baryon and dark matter can suppress the star formation in small mass halos. These effects might be more important if we employ even fainter galaxies in the MCMC. We use an exponential form for the duty cycle to take account of the suppression of star formation in small galaxies due to supernova feedback following \cite{Park2019MNRAS.484..933P}. The exponential form fits the galaxy evolution in hydrodynamical numerical simulations. Although the model is suggested from numerical simulations and it reproduces the LFs well, there are uncertainties due to the parametrization and other astrophysical effects that might impact on the constraints on the PPS.

 The SFR is calculated using Eq.~\eqref{eq:SFR_Mhalo}, indicating that $f_{*,10}$ and $t_{\rm star}$ are degenerated perfectly. Since we use liner- and log-uniform priors for $t_{\rm star}$ and $f_{*,10}$, respectively, the degeneracy at $t_{\rm star}<0.1$ might not be sampled well in our analysis. The degeneracy can be resolved by adding constrains on the star formation rate and stellar mass relation, the so-called main sequence (MS) relation. In \cite{2017ApJ...847...76S}, for example, the MS relation of galaxies at $5<z<6$ has been reported. By comparing their results and Eq.~\eqref{eq:SFR_Mhalo}, we find that $t_{\rm star}\approx 0.1$ is appropriate. On the other hand, in \cite{2020arXiv200600013B}, the lower SFR for galaxies at $z>6$ has been reported, which indicates that $t_{\rm star}\approx 0.5$ might be preferred. We emphasize that this degeneracy does not affect our upper limits on PPS since $t_{\rm star}$ and $f_{*,10}$ do not correlate with PPS parameters.

The prior range of parameters can affect the constraints. Although $\alpha_{\rm star}=0.5$ is preferred as shown in \cite{Park2019MNRAS.484..933P} and numerical simulations indicate the $\alpha_{\rm star} \approx 1$ (e.g., \cite{2017MNRAS.470.2791C}), the prior range can be extend to $\alpha_{\rm star}>1$. In this case, PPS parameters can be degenerate with $\alpha_{\rm star}$ outside of our prior range at $z=8$ and $z=9$ although the constraints should be dominated by the contribution from LFs at $z=6$ and $z=7$. Thus, by extending the prior range, the constraints might become weaker but this effect should be marginal. The range of the other parameters are well motivated and the effects of the prior range on the constraints should be negligible. The model has no physical meaning for $f_{\rm *,10}>1$, $t_{\rm star}>1$, and $t_{\rm star}<0$. The lower and upper bounds of $M_{\rm turn}$ originate from atomic cooling and the faint end of observed LFs \cite{Park2019MNRAS.484..933P}.

%%%%%%%%%%%%%%%%%%%%%%%%%%%%%%%%%%%%%%%%%%%%%%%%%%%%%%%%%%%%%%%%%%
%%%%%%%%%%%%%%%%%%%%%%%%%%%%%%%%%%%%%%%%%%%%%%%%%%%%%%%%%%%%%%%%%%
\section{Conclusion and Discussions \label{sec:conclusion} }

We have derived constraints on small scale primordial power spectrum (PPS) by using high-redshift galaxy UV luminosity functions. Since the halo mass function at low masses has contribution from small scale PPS, we can constrain PPS at small scales from the observed number density of faint galaxies.

We have employed observed UV LFs at $z=6$--$10$ from Hubble Frontier Fields and performed the MCMC analysis to constrain model parameters. For the assumption of astrophysics, we have followed the model used in \cite{Park2019MNRAS.484..933P}. We have considered 5 different parametrizations to model PPS at small scales.

We have found that the running parameters $\alpha_s$ and $\beta_s$ for Parametrization I are constrained by the UV LFs although the constraints are weaker than those obtained by Planck.

We have also considered models with binned PPS at $k>10~\rm Mpc^{-1}$ (Parametrization II--IV) assuming the independent amplitude for each bin. For Parametrization III, 95\% upper limits are obtained as $\log q_1<0.15$ and $\log q_2<3.2$. These constraints translate into ${\cal P}_{\zeta} (10<k<100~{\rm Mpc}^{-1}) < {\cal O}(10^{-8})$ and ${\cal P}_{\zeta} (100<k<1000~{\rm Mpc}^{-1}) < {\cal O}(10^{-5})$. 
% The upper limit at $10<k<100~ \rm Mpc^{-1}$ is tighter than previous constraints and the upper limit at $100<k<1000~ \rm Mpc^{-1}$ is comparable to previous bounds (e.g., \cite{Bringmann:2011ut}).

Here we compare our constraint with other small scale probes of PPS. Lyman-$\alpha$ forest data  constrains  PPS as ${\cal P}_\zeta  \sim (1 - 3.5) \times 10^{-9}$ at the scale of $k \sim {\cal O}(1)~{\rm Mpc}^{-1}$ \cite{2011MNRAS.413.1717B}, which is very severe although the scales probed is relatively limited.  Primordial black holes can also constrain the small scale amplitude for a broad range of scale as down to  $k \sim {\cal O}(10^{20})~{\rm Mpc}^{-1}$, however, its upper bound is ${\cal P}_\zeta < {\cal O}(10^{-1}) - {\cal O}(10^{-2})$  \cite{Josan:2009qn,Sato-Polito:2019hws}, which is rather weak.  CMB spectral $\mu$ distortion can also give a bound on PPS at small scales. From COBE/FIRAS data, one can limit PPS as ${\cal P}_\zeta ( 10 < k < 10^4~{\rm Mpc}^{-1}) < {\cal O}(10^{-5})$ \cite{Chluba:2012we,Chluba:2019kpb}. Ultracompact minihalo (UCMHs) can also put bound on PPS at small scales as ${\cal P}_\zeta (10 < k < 10^7~{\rm Mpc}^{-1}) < {\cal O}( 10^{-6})  -{\cal O}( 10^{-7})$ \cite{Bringmann:2011ut}, which is relatively severe although it depends on the nature of dark matter. Compared to these other probes, our constraints from galaxy UV luminosity function can put a severe constraint around the scale of $10 < k < 1000~{\rm Mpc}^{-1}$. Although astrophysics modeling may give some uncertainties to the bound, it would not change the constraints by many orders of magnitude and such uncertainties would be reduced by understanding astrophysical aspects more with  other observations, e.g., future 21 cm signal at high redshifts \cite{Park2019MNRAS.484..933P}. Therefore our method is complimentary to other probes and, for the scales of $10 < k < 1000~{\rm Mpc}^{-1}$, it can serve as a unique probe of PSS on such small scales.

We have also found that the constraints depend on the filter used to calculate the mass variance $\sigma$. Since the appropriate filter for the enhanced PPS model is unknown, we have compared results with smooth-$k$, top-hat, and Gaussian filters to estimate uncertainties associated with the filter. We have found that the constraints on the PPS depend on the response of the filter at small scales. For the top-hat filter, the constraints are loosen slightly compared to the results with the smooth-$k$ filter. For the Gaussian filter, the PPS at small scales cannot contribute to the mass variance, and as a result PPS at $k>100~\rm Mpc^{-1}$ cannot be constrained. Even for the Gaussian filter, however, $q_1$ is constrained to $q_1<10^{1.2}$. 

We have also used the observed stellar to halo mass ratio (SHMR) in our MCMC analysis. The constraints from SHMR help tighten constraints on the astrophysical parameters, which lead to improved constraints on the PPS parameters in some cases.

We have also investigated the effects of cosmological assumption on the astrophysical parameters estimation from the UV LFs and found that $\alpha_{\rm star}$ and $f_{*,10}$ can degenerate with the PPS parameters, which indicates that the assumption on cosmology affects the study of astrophysics and vice versa. 

Our study shows that astrophysical data such as UV LFs can be a useful probe of cosmology. On the other hand, when one investigates astrophysical aspects, it would also be necessary to take account of the uncertainty of the assumption on cosmological models. Although it looks a bit cumbersome to consider both astrophysical and cosmological aspects, it could also lead to interesting avenue of research ahead of future observations such as James Webb Space Telescope.

%%%%%%%%%%%%%%%%%%%%%%%%%%%%%%%%%%%%%%%%%%%%%%%%%%%%%%%%%%%%%%%%%%
%%%%%%%%%%%%%%%%%%%%%%%%%%%%%%%%%%%%%%%%%%%%%%%%%%%%%%%%%%%%%%%%%%
\begin{acknowledgments}
We thank K. Shimasaku and M. Ishigaki for useful discussions. 
This work is partially supported by JSPS KAKENHI Grant Number 18K03693 (MO), 17H01131 (TT), 19K03874 (TT), 16H05999 (KT), and 20H00180 (KT), MEXT KAKENHI Grant Number 19H05110 (TT), and Bilateral Joint Research Projects of JSPS (KT). SY is supported by JSPS Overseas Research Fellowships. In this work, we use packages pygtc \cite{Bocquet2016} and emcee \cite{emcee}.

% We wish to acknowledge the support of the author community in using
% REV\TeX{}, offering suggestions and encouragement, testing new versions,
% \dots.
\end{acknowledgments}

\bigskip
\appendix

\section{\label{sec:Amv} Mass Variance}

\begin{figure*}[!t]
\includegraphics[width=7.5cm]{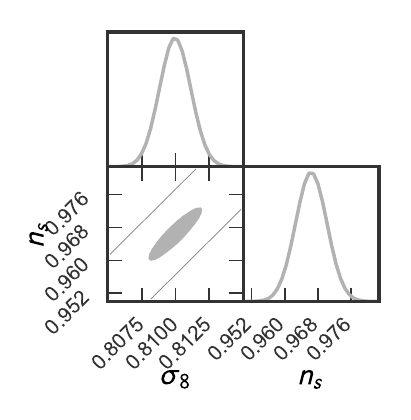}
\caption{\label{fig:sigma8ns} For Parametrization III with the smooth-k filter, the result of MCMC using all UV LFs and the $n_s$ prior. The $\sigma_8$ distribution is calculated from all samples. }
\end{figure*}

Since the amplitude of PPS is fixed and the $n_s$ and $q_{i}$ ($i$=1, 2, 3) are parameters used in MCMC, the value of $\sigma_8$ might break its constraints. To ensure that our model predicts $\sigma_8$ that is consistent with other results, we convert the MCMC samples to $\sigma_8$ as shown in Fig.~\ref{fig:sigma8ns}. 
Our result is consistent with e.g., $0.8102 \pm 0.0060$ in \cite{Akrami:2018odb} well within 1$\sigma$ error.

\section{\label{sec:Afr} Full Results}

\begin{figure*}[!t]
\includegraphics[width=10.5cm]{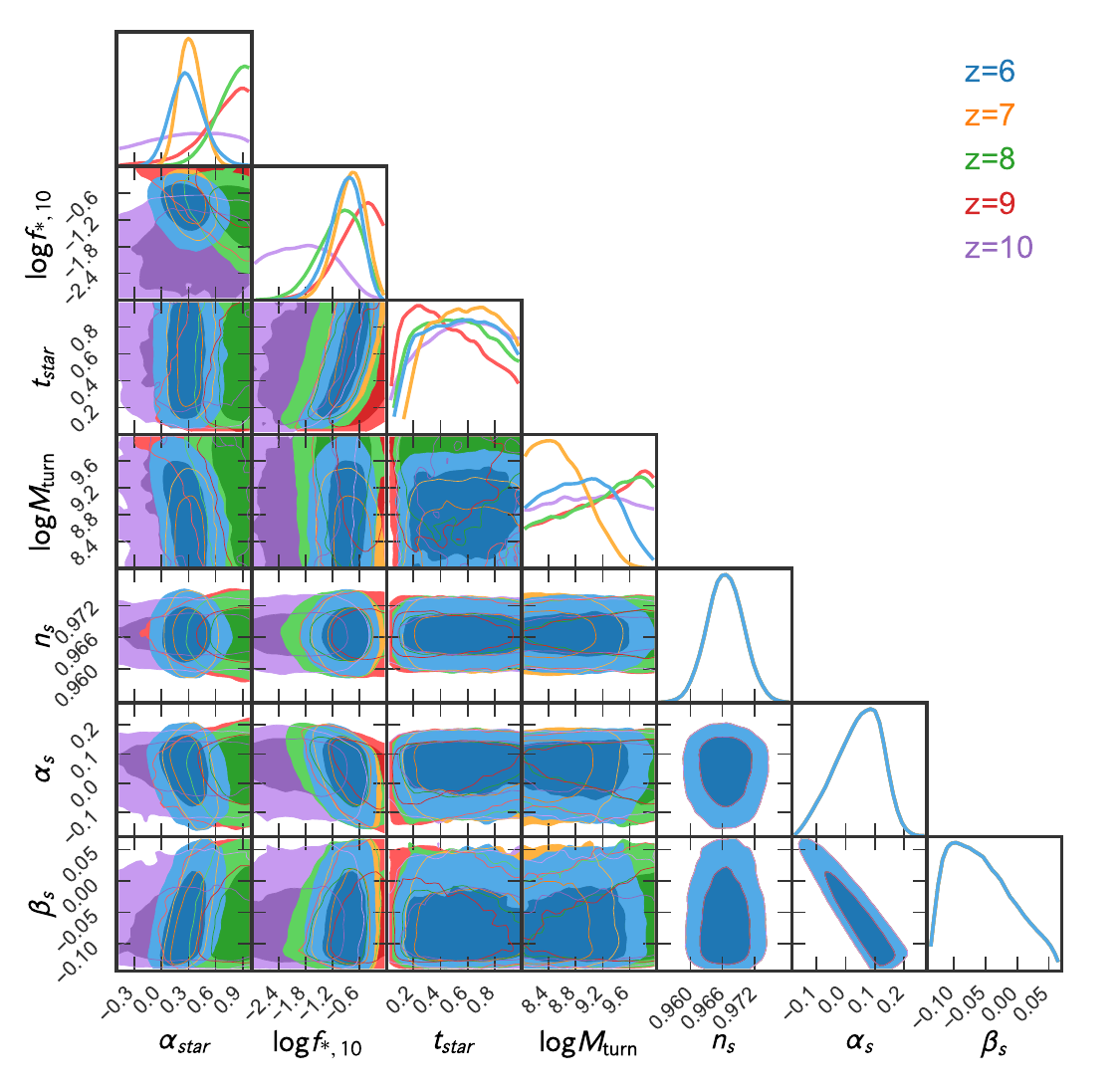}
\caption{\label{fig:mcmcMC1} Result of MCMC using 24 parameters including astrophysical parameters in individual redshift bins (4 parameters $\times$ 5 redshift bins = 20 parameters), spectral index $n_s$ and PPS parameters. This result is for Parametrization I and discussed in Sec.~\ref{sec:const_param_I}. For the constraints, we use UV LFs at all redshifts and Planck constraint on $n_s$ simultaneously.}
\end{figure*}

\begin{figure*}[!t]
\includegraphics[width=10.5cm]{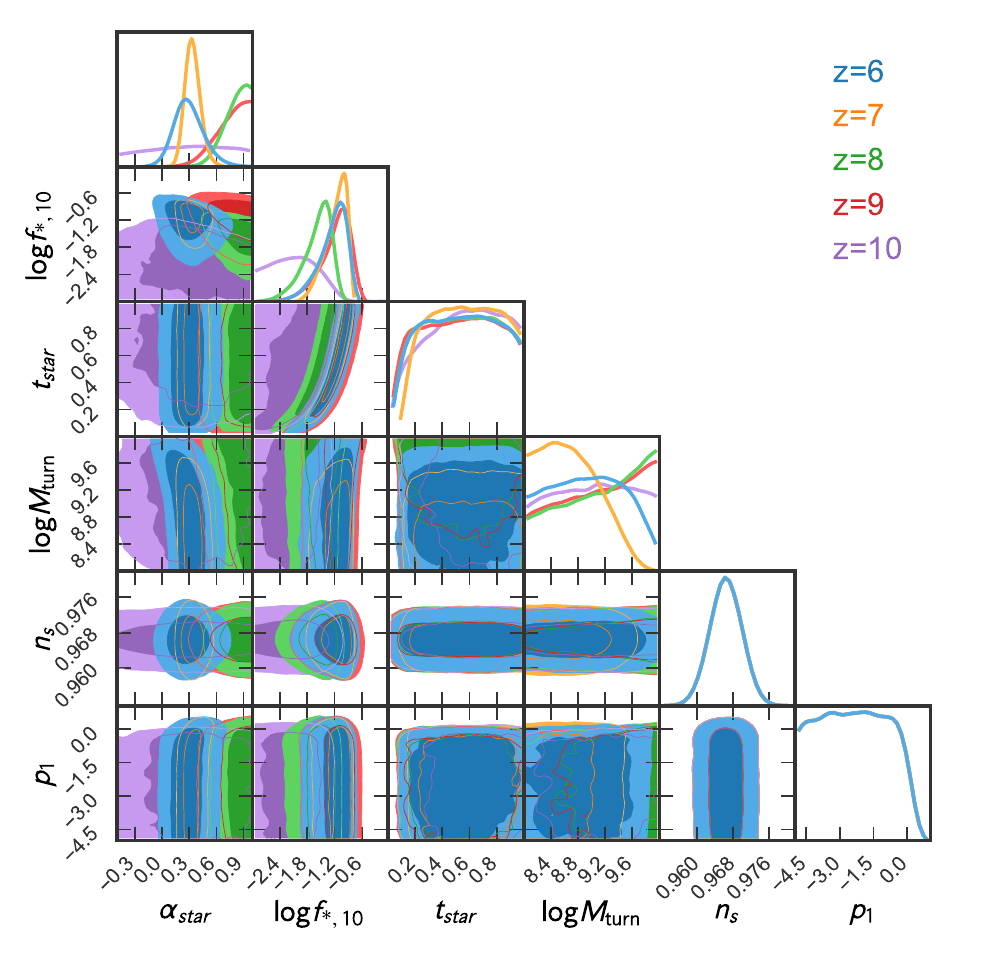}
\caption{\label{fig:mcmcMC0} Same as Fig.~\ref{fig:mcmcMC1}, but for Parametrization II. This result is discussed in Sec.~\ref{sec:const_param_II}. }
\end{figure*}

\begin{figure*}[!t]
\includegraphics[width=10.5cm]{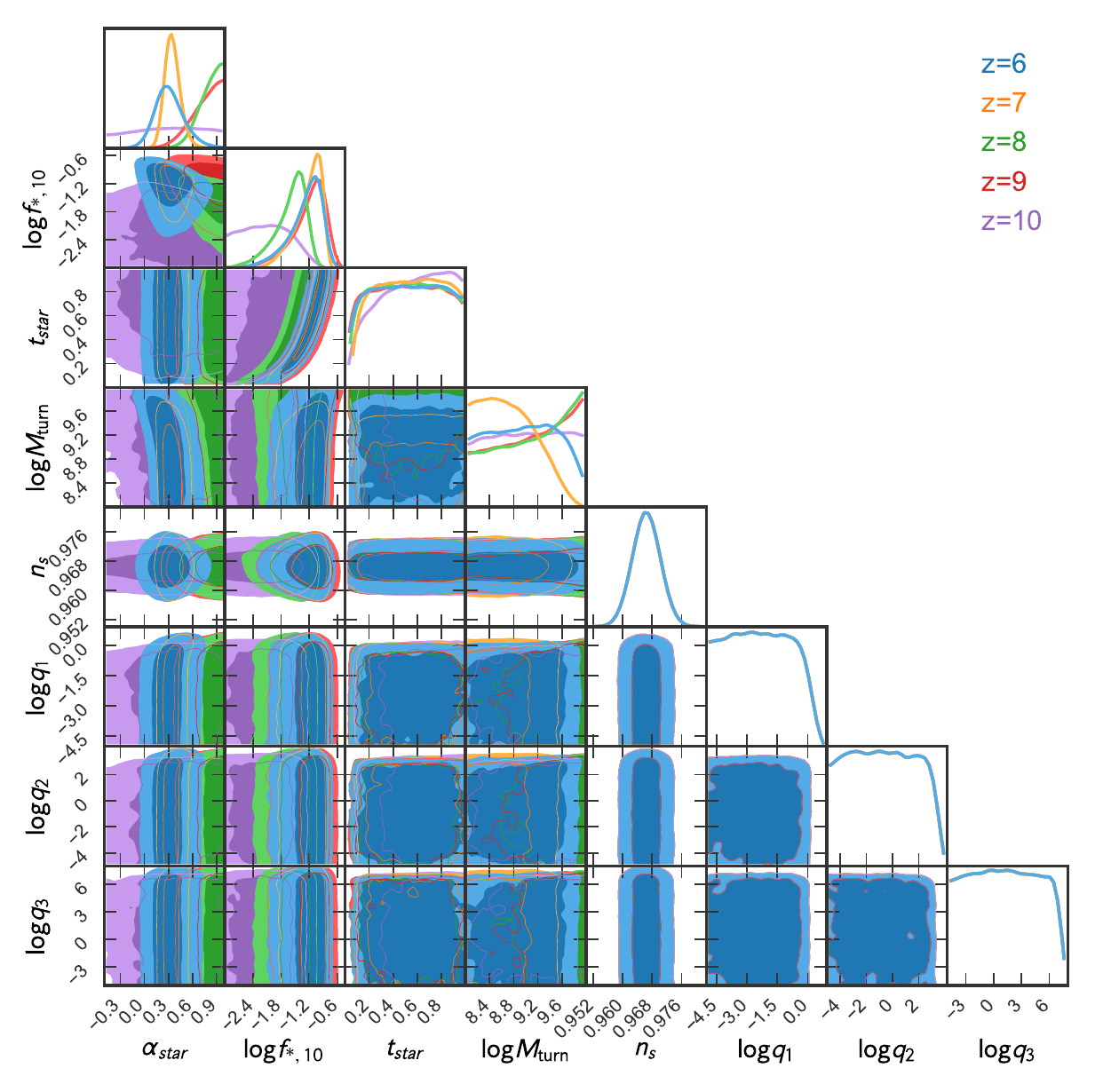}
\caption{\label{fig:mcmcMC2} Same as Fig.~\ref{fig:mcmcMC1}, but for Parametrization III. This result is discussed in Sec.~\ref{sec:const_param_III}, and best fit astrophysical parameters and upper limits on $q_i$ ($i$=1, 2, 3) are listed in Table~\ref{tab:mcmcz} and \ref{tab:mcmcpps} respectively. }
\end{figure*}

\begin{figure*}[!t]
\includegraphics[width=10.5cm]{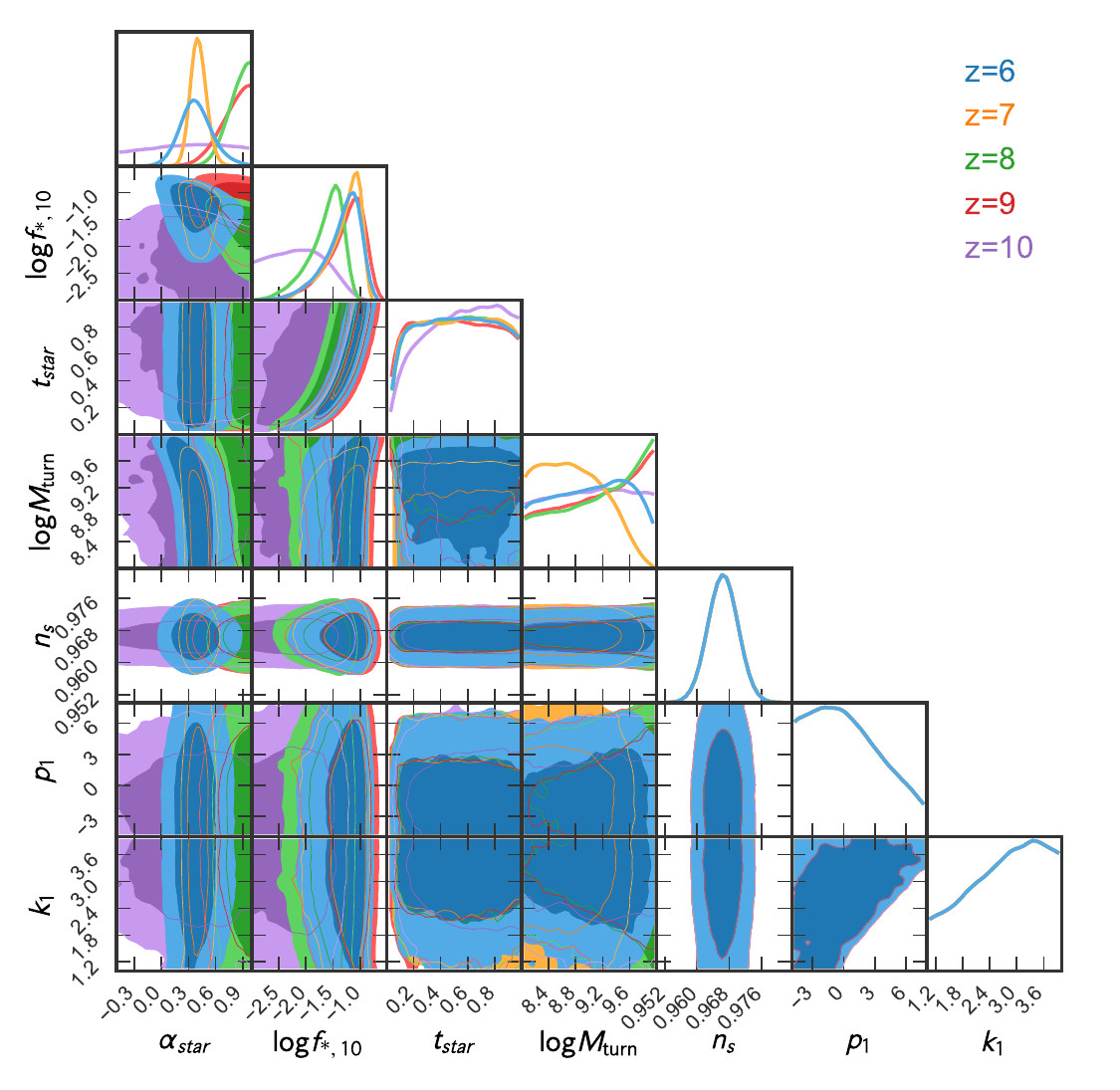}
\caption{\label{fig:mcmcMC3} Same as Fig.~\ref{fig:mcmcMC1}, but for Parametrization IV. This result is discussed in Sec.~\ref{sec:const_param_IV}. This result is discussed in Sec.~\ref{sec:const_param_IV}.}
\end{figure*}

\begin{figure*}[!t]
\includegraphics[width=10.5cm]{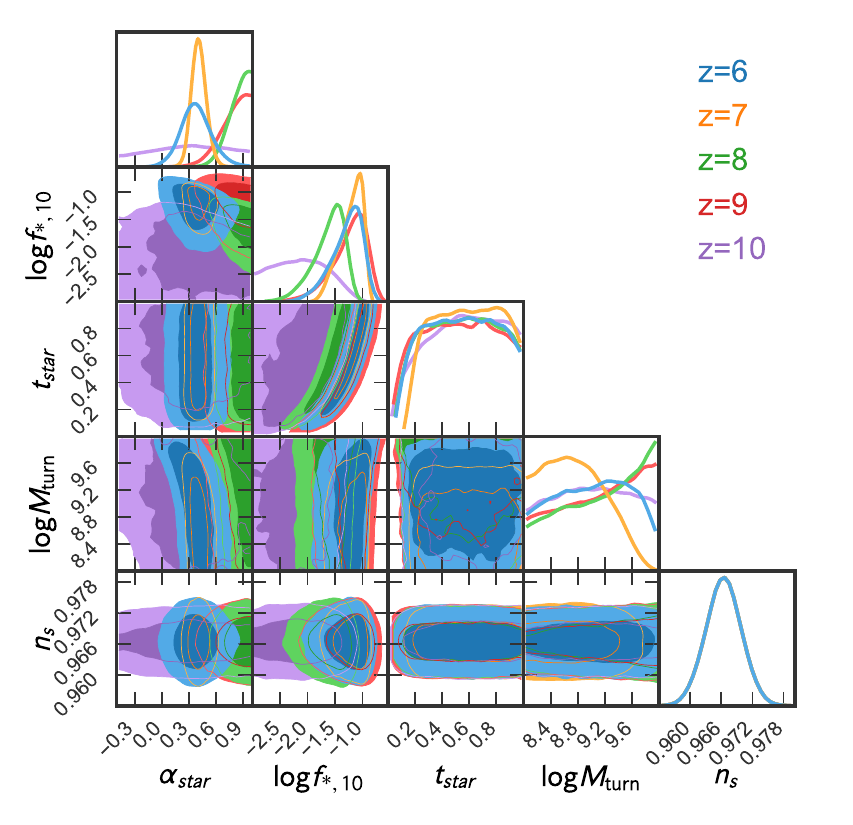}
\caption{\label{fig:mcmcMC5}Same as Fig.~\ref{fig:mcmcMC1}, but for Parametrization V. This result is discussed in Sec.~\ref{sec:const_param_V}.}
\end{figure*}

\begin{figure*}[!t]
\includegraphics[width=10.5cm]{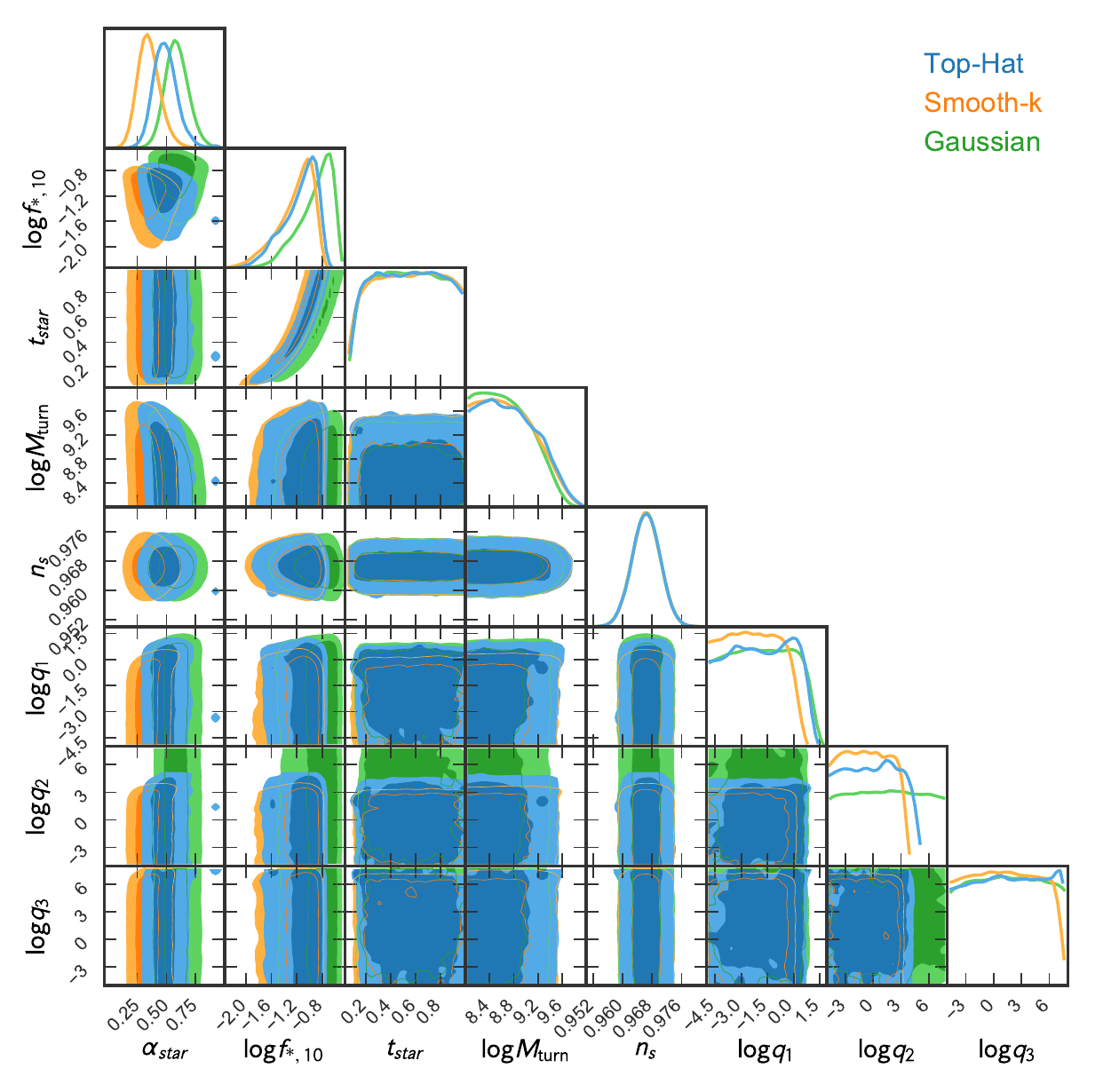}
\caption{\label{fig:mcmcfilter} Result of MCMC using all LFs and Planck prior for various filters. The astrophysical parameter at $z=7$ is shown for reference. This result is discussed in Sec.~\ref{sec:const_param_III}. }
\end{figure*}

We show the full results of the MCMC analysis using all UV LFs and $n_s$ prior. Fig.~\ref{fig:mcmcMC1}, \ref{fig:mcmcMC0}, \ref{fig:mcmcMC2}, \ref{fig:mcmcMC3}, and \ref{fig:mcmcMC5} represent the results of Parametrization I, II, III, IV, and V, respectively. 
Fig.~ \ref{fig:mcmcfilter} is the result for Parametrization III but comparing the results with different filters. 

%\label{fig:mcmcall} Result of MCMC using 24 parameters including redshift independent astrophysical parameters, spectral index $n_s$ and three PPS parameters.

% \begin{figure*}[!t]
% \includegraphics[width=15.5cm]{figure/model_MC2_WF0_all.pdf}
% \caption{\label{fig:mcmc}}
% \end{figure*}

% \begin{figure*}[!t]
% \includegraphics[width=15.5cm]{figure/model_MC2_WF2_all.pdf}
% \caption{\label{fig:mcmc}}
% \end{figure*}

\nocite{*}

\bibliography{bibfile}% Produces the bibliography via BibTeX.

\end{document}